\documentclass[journal]{IEEEtran}
\IEEEoverridecommandlockouts

\usepackage{booktabs} 
\usepackage{amsmath}
\usepackage{siunitx}
\usepackage{amsthm}
\usepackage{relsize}
\usepackage{amssymb}
\usepackage{algorithm}
\usepackage{booktabs}
\usepackage{verbatim} 
\usepackage{subfigure}
\usepackage{balance}
\usepackage{graphicx}
\usepackage{tikz}
\usepackage{algpseudocode}
\usepackage[bottom]{footmisc}
\usetikzlibrary{fit,positioning}
\usepackage[utf8]{inputenc}
\usepackage[english]{babel}
\usepackage{enumitem} 
\usepackage{url}
\usepackage{caption}
\usepackage{tikz}
\usepackage{textcomp}

\newcommand{\dz}[1]{\textcolor{black}{#1}}
\newcommand{\dznew}[1]{\textcolor{black}{#1}}
\newcommand{\dzf}[1]{\textcolor{black}{#1}}
\newcommand{\rv}[1]{\textcolor{black}{#1}}

\newcommand{\zy}[1]{\textcolor{black}{#1}}

\ifCLASSOPTIONcompsoc
  \usepackage[nocompress]{cite}
\else
  \usepackage{cite}
\fi

\ifCLASSINFOpdf
\else
\fi

\newcommand\copyrighttext{
  \footnotesize \textcopyright 2020 IEEE. Personal use of this material is permitted.
  Permission from IEEE must be obtained for all other uses, in any current or future
  media, including reprinting/republishing this material for advertising or promotional
  purposes, creating new collective works, for resale or redistribution to servers or
  lists, or reuse of any copyrighted component of this work in other works.}
\newcommand\copyrightnotice{
\begin{tikzpicture}[remember picture,overlay]
\node[anchor=south,yshift=7pt] at (current page.south) {\fbox{\parbox{\dimexpr\textwidth-\fboxsep-\fboxrule\relax}{\copyrighttext}}};
\end{tikzpicture}
}

\begin{document}

\title{Towards Privacy-aware Task Allocation in Social Sensing based Edge Computing Systems}

%
%
%
%
\author{Daniel (Yue) Zhang,~\IEEEmembership{Student Member,~IEEE,} Yue Ma,~\IEEEmembership{Student Member,~IEEE,} X. Sharon Hu,~\IEEEmembership{Fellow,~IEEE,}  and Dong Wang,~\IEEEmembership{Member,~IEEE}
\IEEEcompsocitemizethanks{\IEEEcompsocthanksitem The authors are with the Department of Computer Science and Engineering
, University of Notre Dame, Notre Dame, IN 46556.\protect\\
E-mail: \{yzhang40, yma1, shu, dwang5\}@nd.edu}\\
}

\maketitle

\copyrightnotice


%
\begin{abstract}
With the advance in mobile computing, Internet of Things, and ubiquitous wireless connectivity, social sensing based edge computing (SSEC) has emerged as a new computation paradigm where people and their personally owned devices collect sensor measurements from the physical world and process them at the edge of the network. This paper focuses on a privacy-aware task allocation problem where the goal is to optimize the computation task allocation in SSEC systems while respecting the users' \emph{customized} privacy settings. 
It introduces a novel Game-theoretic Privacy-aware Task Allocation (G-PATA) framework to achieve the goal. G-PATA includes (i) a \emph{bottom-up} game-theoretic model to generate the maximum payoffs at end devices while satisfying the end user's privacy settings; (ii) a \emph{top-down} incentive  scheme to adjust the rewards for the tasks to ensure that the task allocation decisions made by end devices meet the Quality of Service (QoS) requirements of the applications. Furthermore, the framework incorporates an efficient load balancing and iteration reduction component to adapt to the dynamic changes in status and privacy configurations of end devices. The G-PATA framework was implemented on a real-world edge computing platform that consists of heterogeneous end devices (Jetson TX1 and TK1 boards, and Raspberry Pi3). We compare G-PATA with state-of-the-art task allocation schemes through two real-world social sensing applications. The results show that G-PATA significantly outperforms existing approaches under various privacy settings (our scheme achieved as much as 47\% improvements in delay reduction for the application and 15\% more payoffs for end devices compared to the baselines.).   
\end{abstract}

\IEEEpeerreviewmaketitle

	\section{Introduction}\label{sec:intro}
\IEEEPARstart{W}{ith} ubiquitous Internet connectivity and the proliferation of portable devices, social sensing has become a new sensing paradigm for collecting real-time measurements about the physical world from humans or mobile devices on their behalf~\cite{wang2015social}. Examples of social sensing applications include obtaining real-time situation awareness in the aftermath of a disaster using self-reported observations from citizens~\cite{danielrtas}, monitoring the air quality of a city using inputs from people with portable sensors~\cite{artikis2014heterogeneous}, and detecting real-time traffic congestion using mobile phone apps of drivers~\cite{chen2012crowdsourcing}.
Recent development of edge computing pushes the frontier of computation, service, and data collection 
to the edge of the network~\cite{ahmed2016survey,hu2015mobile}, and brings new opportunities for social sensing applications. By adopting edge computing, social sensing applications can offload both the sensing and computational tasks to \emph{privately-owned} smart devices (e.g., smart phones and nodes in the Internet of Things). We refer to this new application paradigm as Social Sensing based Edge Computing (SSEC)  \cite{danielrtas, wang2019social,zhang2019social, zhang2019edgebatch}.

\dznew{One key concern in SSEC is \emph{privacy} where the end users' private information \dzf{(e.g., location and device statuses such as CPU usage and frequency)} may be leaked to the server or other participants of the application. Such privacy concern originates from the diversified ownership in SSEC system where the devices are owned by individuals with diverse interests and affiliations. Therefore it is very difficult to form the mutual trust among these participants in the system. Furthermore, it  may cause serious consequences by sharing the private information with the server  (e.g., identity theft and information harvesting should a malicious attacker compromises the server or  the server is malicious by itself) \cite{nathanicpads}.}

In light of the unique privacy concern of SSEC, we study a critical problem of  \emph{privacy-aware task allocation}  in the SSEC system. Our goal is to effectively allocate the computation tasks in an application to available resource in the system (e.g., end devices) to meet the Quality of Service (QoS) requirements of the application (e.g., deadline hit rate, end-to-end delay) while respecting the  privacy constraints from the end users. \rv{Different from existing literature on privacy-preserving edge computing that assumes uniformed privacy requirements from all users \cite{lu2017lightweight,nathanicpads}, we assume the privacy setting is user-dependent, i.e., each individual end user can have his or her own privacy preference.  Compared to the flat security and privacy protection policy that only enforces an ``on-off" mode, a customized policy would treat users based on their unique privacy preferences. The key advantage of such personalized privacy design is that the server can benefit from device owners who are not paranoid to share their private information. We argue that such customized design is \emph{practical}as it is well observed that privacy is both context and user-dependent \cite{zhang2014cross, zhou2015information,xiao2006personalized,wang2018personalized}.}

\dz{We found the privacy-aware task allocation problem to be uniquely challenging due to the complex tradeoff between privacy settings of the end users (i.e., owners of end devices) and the optimization of task allocation in the system.  In particular, in order to obtain a globally optimal task allocation scheme, the SSEC system usually needs to collect the location and status information of all end devices \cite{habak2015femto}. During this process, privacy is an important concern in SSEC for multiple reasons.} First, the sensing data collected from end devices are often privacy sensitive. For example, in a social sensing application of destruction detection in the aftermath of a disaster~\cite{zhang2019crowdlearn}, it's important to associate the sensing report with the accurate location of the report. However, the reported location could potentially reveal the reporter's residence information. Second, both servers and end devices are not fully trusted by one another in SSEC and may intentionally try to harvest other users' private information for either revenue  or malicious intents \cite{cornelius2008anonysense}. \dz{Last but not least, a malicious server can leverage side channels (such as the task preferences of the users \cite{li2017scalable}) to infer the private information without directly accessing it from the end users. 
}   \dznew{ With the above challenges in mind, we focus on two important aspects unique to this privacy-aware task allocation in  SSEC systems in this paper}: \emph{rational end devices} and \emph{partial information}. 

\emph{Rational end devices:} \rv{One unique design of SSEC is to leverage privately owned end devices at the edge, rather than merely using dedicated nodes (edge/fog servers such as micro data centers and cloudlets) to perform sensing and computational tasks. The owners of these end devices are rational actors and often have inconsistent or even conflicting objectives with the application.} From the application's perspective, it is of its best interest to obtain better knowledge on the status of each end device to optimize the task allocation performance. For example, the location information of an end device will be helpful for the application to decide where to offload the computation tasks. However, the end devices' owners may not be willing to share such information due to their privacy concerns. This is in sharp contrast to traditional distributed computing systems where computational resources are fully cooperative and directly controlled by the application. Several privacy-preserving schemes have been developed to allow participants to hide their private information via anonymity techniques \cite{li2018scalable,nathanicpads,cornelius2008anonysense}. However, these solutions did not explicitly consider the performance trade-off between the application performance and the privacy requirements of the end users. The rational nature of end devices in SSEC must be carefully addressed by developing a new task allocation model that respects the discrepancies between the objectives of applications and end devices.

\emph{Partial Information:}  The privacy configurations of end devices in SSEC often lead to partial information of the devices to be shared, which prevents the system from achieving the optimal task allocation decision. For example, if most of the end devices in the system are unwilling to reveal their computing power information (e.g., CPU frequency and CPU usage), the server will have little knowledge on which end device is more suitable for executing a computationally intensive task.  The problem becomes more challenging in SSEC system where end devices (end users) often have different privacy preferences (e.g., some end users are willing to expose their location information where others may not), leading to various degrees of partial information in the task allocation process. With such incomplete knowledge of the system, not only the server cannot make optimal decisions, but also the end devices cannot accurately conjecture the behaviors of other end devices and decide their own best strategies in obtaining the computation tasks.  This user-defined partial information challenge is unique in the privacy-aware SSEC system and cannot be directly solved by current task allocation schemes that either assume full information disclosure or ignore the self-controlled privacy level defined by end users.

To address the above challenges, we develop a novel Game-theoretic Privacy-aware Task Allocation (G-PATA) framework. To address the rational end devices challenge, we develop a bottom-up task allocation model based on game theory to incentivize end devices to pick the tasks that provide them maximum payoffs. A dynamic reward mechanism is designed to ensure the QoS of the application is satisfied. To overcome the partial information challenge, we design a privacy-aware negotiation process that allows the server and end devices to jointly make optimal decisions given only limited information on the status of end devices. 

We also implemented a prototype of G-PATA using a real-world SSEC system that consists of Nvidia Jetson TX1 and TK1 boards, and Raspberry Pi3. We evaluated the performance of G-PATA using two real-world delay-sensitive social sensing applications: \emph{Collaborative Traffic Monitoring} \cite{chatzimilioudis2012crowdsourcing} and \emph{Abnormal Event Detection} \cite{mahadevan2010anomaly}. We compared G-PATA with the state-of-the-art task allocation schemes used in edge computing systems. The results show that our scheme achieves a significant performance gain in terms of meeting the performance objectives of both applications and end devices under various privacy settings.

	\section{Motivation and Related Work}
\label{sec:related}

\begin{table*}[htp!]
\centering
\caption{\rv{Summary of Existing Related Work Compared to The  Proposed Work}}
\rv{\label{tab:compare}
\begin{tabular}{|c|c|c|c|c|c|}
\hline
Compared Schemes & \textbf{Customized Privacy}  & \textbf{Rational Edge} & \textbf{Incomplete Information} & \textbf{Optimize Delay} & \textbf{Optimize Energy}   \\ \hline
Our scheme (G-PATA)      & \checkmark  & \checkmark & \checkmark & \checkmark & \checkmark \\ \hline
 Top-down allocation (\hspace{1sp}\cite{zhang2017towards, bertuccelli2009real,zhu2012optimization,su2013elastic,ahmad2008using})      &  &  &  & \checkmark & \checkmark \\ \hline
Auction game (\hspace{1sp}\cite{danielsec, feng2014trac, liu2017decentralized})      &    &  \checkmark & \checkmark&  &  \\ \hline
Congestion game (\hspace{1sp}\cite{danielrtas, danielinfocom} ) &   & \checkmark &  \checkmark & \checkmark & \checkmark \\ \hline
Group bidding (\hspace{1sp}\cite{li2018scalable, li2017scalable}) & & \checkmark &  \checkmark &  &  \\ \hline
Differential privacy (\hspace{1sp}\cite{andres2013geo,he2017customized,jorgensen2015conservative})      & \checkmark    & \checkmark &  \checkmark&  &  \\ \hline
\end{tabular}}
\end{table*}

\subsection{Social Sensing and Edge Computing}
Social sensing has received a significant amount of attention due to the proliferation of low-cost mobile sensors and the ubiquitous Internet connectivity~\cite{R1,rashid2019sead,wang2019age}. Examples of such applications include intelligent transportation systems~\cite{chen2012crowdsourcing},  urban sensing~\cite{zhang2019transland}, and disaster and emergency response~\cite{rashid2019socialcar}. \dz{ Traditional social sensing applications push all the computation tasks to the backend servers, which can be quite ineffective, especially for time-critical applications, due to the limited network bandwidth and high communication latency \cite{zhang2017towards}.} \dz{The edge computing paradigm complements traditional centralized social sensing solutions by offloading computation tasks to the end devices to reduce communication costs and application latency~\cite{shi2016edge}. We refer to the marriage of social sensing and edge computing as Social Sensing based Edge Computing paradigm or SSEC.  There is much untapped potential for highly responsive social sensing applications at the edge. For example, Zhang \emph{et al.} proposed a game theoretic bottom-up task allocation scheme to allow non-cooperative end devices to perform social sensing tasks by providing them with incentives \cite{danielrtas}. Satyanarayanan \emph{et al.} introduced an intermediate layer (i.e., ``cloudlet") located between the cloud and mobile devices to mitigate latency related issues between end devices and servers \cite{satyanarayanan2009case}. A major knowledge gap in these schemes is that they require the end devices to trust the server to manage their privacy. This trust is often misplaced since even trustworthy servers can also suffer from a data-breach \cite{nathanicpads}. In this work, we propose G-PATA, a framework to efficiently allocate social sensing tasks that satisfy user-defined privacy requirements. }

\subsection{Task Allocation}
\dz{The allocation of computation tasks is the core problem in SSEC and many relevant solutions have been proposed in traditional real-time and distributed systems \cite{zhang2017towards, bertuccelli2009real,zhu2012optimization,su2013elastic,ahmad2008using}.} Both top-down and bottom-up approaches have been developed. In a top-down approach, a centralized decision maker (often an algorithm running on the back-end server)  makes global allocation decisions with the assumption that end devices are fully controlled and cooperative  \cite{zhang2017towards,bertuccelli2009real}.  For example, Zhu \emph{et al.} proposed a Mixed Integer Linear Programming based approach to meet the deadlines and minimize the end-to-end latency in hard real-time systems \cite{zhu2012optimization}. Su \emph{et. al} developed a mixed criticality task model to maximize the number of low-criticality tasks being executed without influencing the timeliness of high-criticality tasks~\cite{su2013elastic}. 
Most of the above schemes depend on the accurate schedulability analysis (e.g., estimating the worst-case execution time (WCET) of assigning a task to a device). Such an approach would fail in privacy-aware SSEC where the end users might refuse to provide necessary information to comply with the centralized task allocation schemes \cite{danielinfocom, zhang2019heteroedge}.  Due to the limitations of centralized task allocation schemes, \dz{a decentralized approach that end devices make task allocation decisions by themselves seem to be a better fit to the privacy-aware SSEC system.} Decentralized decision-making schemes have been developed in mobile crowdsensing, wireless networks, and multi-agent systems,  where decisions are made autonomously by devices interacting with each other. For example, Ahmad \emph{et al.} developed a game-theoretic approach for scheduling tasks on multi-core processors to jointly optimize performance and energy~\cite{ahmad2008using}. 
A few auction game based bottom-up approaches are also developed for distributing crowdsourcing tasks to the mobile users \cite{feng2014trac,liu2017decentralized}. In a more recent work, Zhang \emph{et al.} proposed CoGTA, an edge computing system that allows non-cooperative and heterogeneous end devices to trade tasks and claim rewards \cite{danielsec}. However, the above solutions did not address the privacy constraints imposed by end users, thus cannot be directly applied to our problem.

\subsection{Privacy Preservation}
Privacy preservation has always been a key design objective in edge computing and mobile crowd sensing communities. Providing anonymity is a key privacy protection technique that has been applied to many relevant fields such as mobile crowdsourcing and distributed systems. 
For example, Al-Muhtadi \textit{et al.} developed a protocol called Mist to mask user locations in ubiquitous computing~\cite{al2002routing}. Song \textit{et al.} advocate the use of VPNs combined with fuzzy logic to promote privacy in grid computing~\cite{song2004fuzzy}.  A serious limitation of these techniques is that they prevent the application from identifying the contributors of the computation tasks \cite{ni2017security} and the rewards can not be sent to the contributing end devices that accomplish the computation tasks if anonymity techniques are directly applied in SSEC \cite{nathanicpads}. To address this challenge, Vance \emph{et al.} developed a block-chain based decentralized edge computing framework that allocates incentives to end devices without revealing their identity \cite{nathanicpads}. Li \emph{et al.} developed a secure-bidding scheme for bidding tasks among mobile devices while protecting individual privacy \cite{li2018scalable}.

\rv{A set of recent work on customized differential privacy \cite{he2017customized,jorgensen2015conservative,andres2013geo} in location-based systems (LBS) are also highly relevant to our work. For example, Andr{\'e}s \textit{et al.}  proposed a geo-indistinguishability differential privacy framework  that allows  each  individual user  to define customized privacy levels for a particular radius of spatial region \cite{andres2013geo}. Jorgensen \textit{et al.} developed a  personalized differential privacy (PDP) technique that directly generalizes the classical differential privacy to consider different privacy requirements for different users \cite{jorgensen2015conservative}. While these approaches can also provide user-defined privacy protection, they suffer from several limitations.  First, the differential privacy-based approaches focus on the utility-privacy trade-offs rather than providing strict privacy protection. In contrast, in this paper, our assumption is that the private data must be completely indistinguishable and a user will not sacrifice his/her predefined privacy level by trading it with the data utility. Another limitation lies in the fact that existing work have pretty strong assumptions on the granularity of the privacy setting.  For example, most of them are designed to customize privacy for continuous spatial regions (e.g., characterized by a radius or grids).  These assumptions are often not practical from a usability standpoint in our problem. This is because people may prefer all types of privacy setting options such as hierarchical settings (city/state/country), continuous settings (CPU frequency, memory), or categorical settings (device is sleep/not). To our knowledge, existing differential privacy-based approaches do not provide such flexibility on the privacy settings for users. In contrast, the proposed scheme addresses this problem by modeling the data accessibility under privacy constraints by defining  a probabilistic density function over a specific uncertainty region. These functions are general and can be defined for various types of data.}
\rv{\subsection{Summary}
Finally, we summarize the approaches, advantages, and disadvantages of the related work in Table \ref{tab:compare}. It is clear that the proposed G-PATA scheme is the first one to jointly optimize the competing objectives (energy minimization of the rational edge and delay minimization at the server) while satisfying the customized privacy requirements.}

	\newtheorem{myDef}{DEFINITION}
\section{Problem Formulation} \label{sec:problem}

In this section, we present the privacy-aware task allocation problem for delay-sensitive social sensing applications in edge computing systems. We first present the task model, assumptions, and the privacy model we used, and then formally define the objectives of our problem.  \rv{The definitions of the notations are summarized in Table~\ref{tab:notations}.}

\begin{table}[htb!]

    \caption{\rv{Summary of Notations}}
    \rv{\centering
    \begin{tabular}{|l|l|}
        \hline
        $t$  & The $t^{th}$ sensing cycle, $t\in \{1,2,...,T\}$\\ \hline
        $E_x$  & The $x^{th}$ end device, $x\in \{1,2,...,X\}$\\ \hline
        $S_y$  & The $y^{th}$ edge server, $y\in \{1,2,...,Y\}$\\ \hline
         $\tau_m$  & The $m^{th}$ task, $m\in \{1,2,...,M\}$\\ \hline
             $\Delta$  & The common deadline of the tasks\\ \hline
             $DI_{x,z}$  & The $z^{th}$ private information of $E_x$, $z\in \{1,2,...,Z\}$\\ \hline
             $\overline{DI}_{x,z}$  & The privacy requirement for $DI_{x,z}$\\ \hline
             $\pi_{m,x}$  & The cost of $E_x$ for performing $\tau_m$\\ \hline
             $\mathcal{D}_m$  & The end-to-end delay of $\tau_m$\\ \hline
             $q_{j,i}$  & The quality score of end device $E_j$ when performing $\tau_i$\\ \hline
             $R_i$  & The reward of a task $\tau_i$\\ \hline
             $u_{j,i}$  &  The payoff of end device $E_j$ for picking $\tau_i$\\ \hline
             $S(j)$  &  The edge server assigned to $E_j$ \\ \hline
    \end{tabular}}
    	\vspace{-0.1in}
    \label{tab:notations}
\end{table}

\subsection{Overview of SSEC Systems}

\begin{figure}[!htb]
	\centering
	\includegraphics[width=0.45\textwidth]{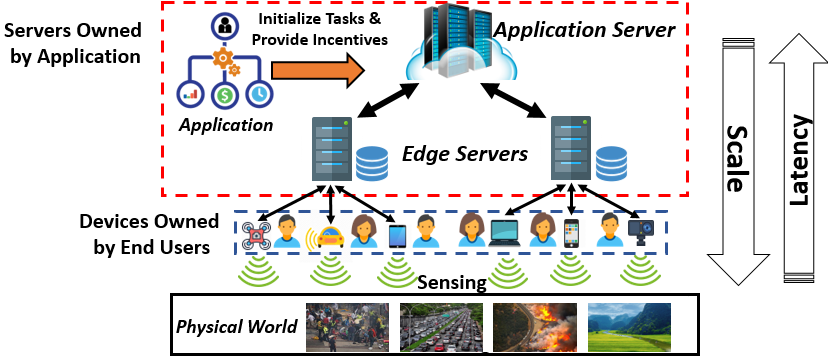}
	\caption{An Overview of The Social Sensing based Edge Computing System Architecture}
	\vspace{-0.1in}
	\label{fig:ecss}
\end{figure}

\dz{A typical SSEC system has three layers (as shown in Figure \ref{fig:ecss}): end devices, edge servers, and a remote application server. Let $E\!D = \{E_1, E_2,...,E_X\}$ denote the set of all end devices in the system.  We assume the social sensing application manages a set of $Y$ edge servers - $E\!S = \{S_1, S_2, ...,S_Y\}$. These edge servers provide local data processing/storage capabilities and coordinate the connected local end devices in accomplishing the allocated tasks.  We assume each end device is assigned to one particular edge server and the result of the computation task finished by an end device should be offloaded to the assigned edge server for further processing. The remote application server $A\!S$ (often built into a remote data center/cloud) provides a global service interface to all users of interest to the social sensing applications.} 

\subsection{Task Model in SSEC}

In an SSEC system, end devices not only collect sensing data but also contribute significantly to social sensing computation tasks. A social sensing computation task processes the raw sensing data using the algorithm assigned by the application. These tasks can be for data preprocessing to compress the raw sensing data or for feature extraction to extract and provide useful features.

We adopt a \emph{frame-based task model}~\cite{allavena2001scheduling} commonly used in the real-time system community where computation tasks are periodically initialized and all tasks have the same period and deadline. We use $\Delta$ to denote the common deadline of all the tasks in an application. $\Delta$ captures the user desired QoS in terms of when the tasks should be finished. For a given application at the $t^{th}$ sensing cycle (i.e., period), we assume that a total of $M$ tasks  are to be processed, i.e., $Tasks = \{\tau_1(t),\tau_2(t), ...,\tau_{M}(t)\}$.  Each computation task is associated with a 4-tuple: $\tau_m(t) = \{ VI_m(t), VO_m(t), alg_m(t), type_m(t)\}$ where $VI_m$ is the data volume to be processed by task $\tau_m(t)$, and $VO_m(t)$ is the expected size of the output. $alg_m(t)$ and $type_m(t)$ denotes the algorithm to process the input sensing data and the data type of the data (i.e., text, image, video). 
For example, a social sensing application can task a set of vehicles to collect and process the data related to road conditions (e.g., video data using dash cameras).

\subsection{Privacy and Threat Models}
\dz{To capture the privacy concerns of end users in SSEC, we assume that the application defines a set of $Z$ categories of private device information: $\{DI_{x,1}, DI_{x,2}, ... DI_{x,Z}\}$, where $DI_{x,z}$ denotes the $z^{th}$ private information item of device $E_x$. An example of $DI_x$ would be (location: 5th Ave, NYC, USA; CPU frequency: 2.4 GHz: CPU usage: 50\%).}

\begin{figure}[!htb]
    \centering
    \includegraphics[width=0.45\textwidth]{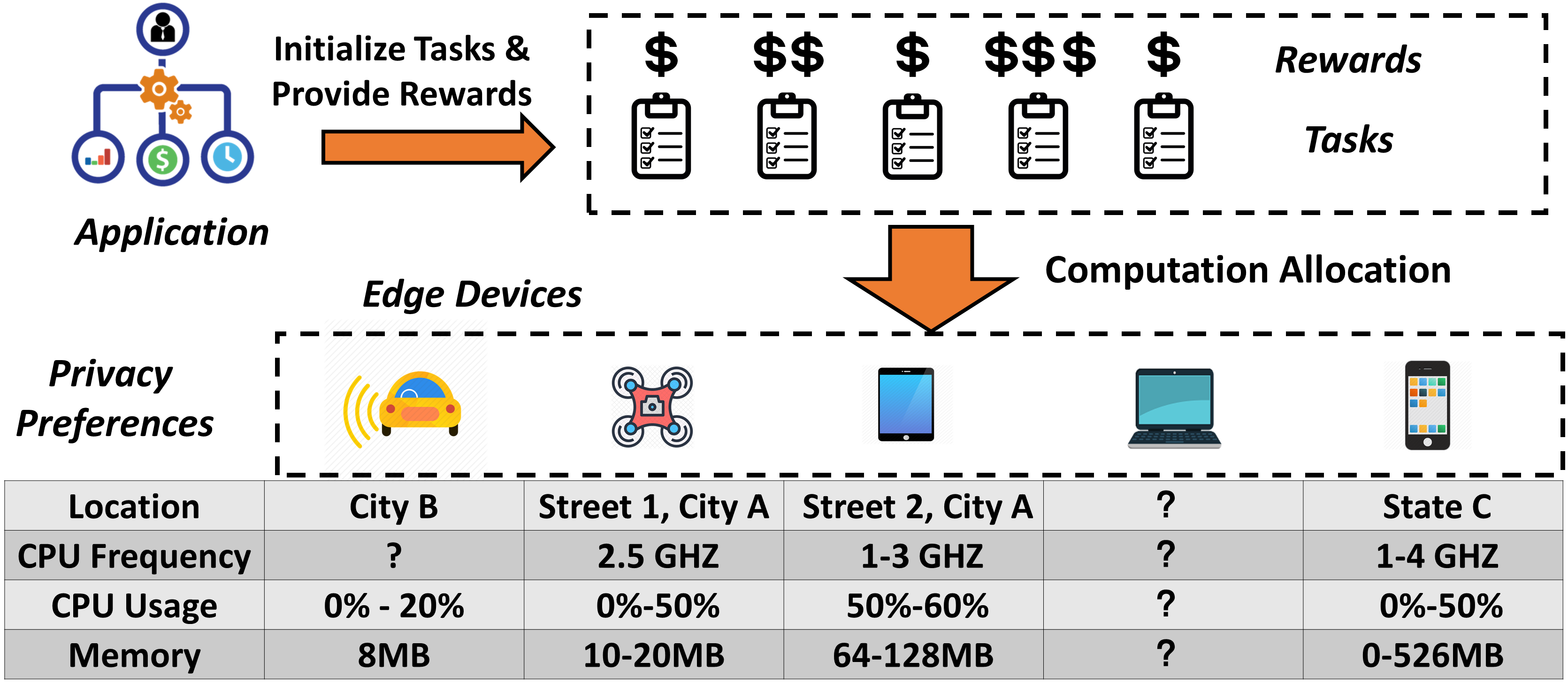}
    \caption{Privacy-aware Task allocation Problem}
    \vspace{-0.1in}
    \label{fig:paca}
\end{figure}

\dz{We assume end users may have different \emph{privacy requirements} for their private  information. The privacy requirements are set up by end users when they first commit to the social sensing application (e.g., by installing a client software) and can be dynamically adjusted during the whole period of the application.  For example, some users are willing to disclose the street address of their locations while others may only be willing to disclose which city they are in. For every end device $E_x$, we define its privacy requirement at the $t^{th}$ sensing cycle as $\{\overline{DI}_{x,1}(t), \overline{DI}_{x,2}(t), ..., \overline{DI}_{x,Z}(t)\}$ where  $\overline{DI}_{x,z}(t)$ denotes the privacy requirement for $DI_{x,z}$ at the $t^{th}$ sensing cycle. We illustrate an example of self-defined privacy requirements in Figure \ref{fig:paca}. In this example, each device has its own privacy requirement. For a particular category of information, some end devices can be very conservative (e.g., only reveal the current CPU frequency with a high uncertainty range such as 1-3 GHz, or even disclose nothing, as denoted as ``?" in the figure) while other devices can be very ``generous'' by sharing the exact CPU frequency information such as 2.5 GHz.} 
\dz{The privacy-aware task allocation problem is to study how to perform optimal task-device mapping given such diversified privacy requirements.}

\rv{We adopt a threat model where the adversaries are \emph{honest-but-curious}, which is a commonly used threat model for crowdsourcing and edge computing related applications \cite{li2018scalable, yang2015security,ni2016secure,abbasinezhad2018design}.  Under such an assumption, the end users, edge servers, as well as the application server  will follow the application protocol truthfully without cheating but they are interested in inferring the private status and location of other end devices.} We then define the \emph{privacy requirement violation} as follows:

\begin{myDef}
    \emph{\dz{Privacy Requirement Violation}:  the privacy requirement of an end device (end user) is violated when the server or any other end devices have access to more private information than the user is willing to disclose.}
\end{myDef}

\dznew{To ensure privacy requirement violation, merely preventing the server or other devices from having direct access to the private information of an end user is not enough. We note that side-channel attacks can obtain a certain amount of private information even the users do not directly disclose such information \cite{li2018scalable}. \zy{For example, in a task allocation framework, if an end device always prefers a task with high computational complexity, then the end device may likely have high computation power  (e.g., low CPU usage and high CPU frequency)~\cite{liu2017decentralized}.} Considering such side-channel based privacy violation, we also define the \emph{decision privacy} as follows:}

\begin{myDef}
\emph{Decision Privacy:   the task allocation decision of each individual end device should not be exposed to other end devices. }
\end{myDef}

\zy{Note that we only protect decision privacy among end devices. The reason is that the server has to know the decisions made by end devices to assign tasks and rewards. In such a context, achieving decision privacy is still particularly challenging in bottom-up approaches, which we elaborate in Section \ref{sec:solution}. }

\rv{We also note that potential information leakage can happen if a malicious server carefully examines the raw sensing data provided by the end devices. One potential attack is via correlated data, where the server can examine the similarity of the data provided by two devices, and infer private information if one device provides more information than the other. Consider a simple example where two devices are reporting air quality index of their current region. Device A is willing to provide full location information to the server while Device B only discloses the current city. The server can then guess the full location of Device B if it reports the exact same air quality index as Device A. Another potential attack is directly extracting private information from the raw data. For example, consider a social sensing application that leverages drones/vehicles to monitor the road conditions using onboard cameras. It is possible to have human faces, car plate numbers, or location identifiers (e.g., street names) in the reported images. This work does not focus on the above privacy leakages related to the raw sensing data. We discuss how our framework can be extended to address this issue in Section  \ref{sec:discuss}.}

\subsection{Reward and Cost Model}
An appropriate reward mechanism is essential to facilitate compliance in SSEC where end devices are not owned by the application. In order to motivate end devices to perform computation tasks, it is often necessary to provide rewards (either monetary or non-monetary) to end users to incentivize their participation in computation tasks and compensate for their costs (e.g., energy) to finish the tasks~\cite{zhang2016incentives}. We define a few terms below.
\begin{myDef}
    \emph{\dz{Edge Cost $\pi_{m,x}$}: we use energy cost to represent the cost that end devices have to pay in completing computation tasks.}
\end{myDef}
\begin{equation}
    \pi_{m,x} = Power_x \cdot c_{m,x}
\end{equation}
$Power_x$ is the average power consumption of end device $E_x$ and is calculated by
\begin{equation}
    Power_x =  Power_{{comp},x} + Power_{{trans},x}
\end{equation}\noindent
where 
$Power_{comp}$ is the power consumption for computation, and  $Power_{trans}$ is the power consumption for data transmission via wireless network and is proportional to the data size been transferred.

\begin{myDef}
    \emph{\dz{Task Rewards}: we assume that the server has a fixed budget for a social sensing application. The server needs to compensate for the end devices' contributions by providing a monetary or non-monetary reward for completing each task (e.g., virtual currency, virtual credit). }
    
\end{myDef}

\begin{myDef}
    \emph{\dz{Edge Payoff $u_x(t)$}: we use $u_x(t)$ to denote the overall benefit that end device $E_x$ receives by executing tasks at the $t^{th}$ sensing cycle. It defines the individual gain (as a function of of cost and reward) of an end device and more details are given in Section \ref{sec:solution}.}
\end{myDef}

\subsection{Problem Statement}
The goal of our problem is to identify an optimized task allocation scheme to best meet the objectives of the end devices and the application without violating the privacy requirements of end users.  The objective of the end devices is to maximize their payoffs, $u_x(t)$, defined in  \dznew{DEFINITION 5}.  To formulate the application objective, we first define a key term \emph{end-to-end delay} as:

\begin{myDef}
     \emph{\dz{End-to-end delay of task($\mathcal{D}_m(t)$)}: the total amount of time taken for the sensing data to be processed by $\tau_m (t)$ and transferred to the local edge server.}
\end{myDef}
  
\dz{Then we use a binary variable $\delta_m (t)$ to represent whether a task misses a deadline or not:}

\begin{equation} \label{eq:deadlinehit}
\begin{split}
 \delta_m (t)= \left\{
\begin{aligned}
&1,~\mathcal{D}_m(t) > \Delta~(deadline~miss)\\
 &0,~otherwise~(deadline~hit)
 \end{aligned}
\right.
 \end{split}
\end{equation}
\noindent

We formally define the application objective as minimizing the deadline miss rate of all tasks.

\dz{Given the above definitions, assumptions and system models, our privacy-aware task allocation problem is a multi-objective constrained optimization problem:}
\begin{equation} \label{eq:dql}
\begin{split}
& \text{maximize} ~~   u_x(t), \forall 1 \le x \le X,  1 \le t \le T \\& \quad\quad\quad\quad  \text{(end device/end user objective)}\\ 
& \text{minimize} ~~   \sum_{m=1}^{M} \delta_m (t),  \forall 1 \le m \le M,  1 \le t \le T \\& \quad\quad\quad\quad \text{(application objective)}\\
& \text{s.t.:}~~ \overline{DI}_{x}(t) \text{is satisfied}, \forall 1 \le x \le X,  1 \le t \le T  \\& \quad\quad\quad\quad \text{(privacy constraints)} \\
\end{split} 
\end{equation} 
\noindent

Note that the task allocation problem for heterogeneous distributed systems is in general NP-hard~\cite{chen2009efficient,zhu2012optimization}. The problem becomes more challenging in this multi-objective formulation where the objectives of the application and end devices are potentially conflicting and additional privacy constraints are imposed. In the next section, we discuss our G-PATA framework to solve this problem.

	\section{A Game-theoretic Privacy-aware  task allocation Framework}
\label{sec:solution}

In this section, we develop a  Game-theoretic Privacy-aware  Task Allocation (G-PATA) framework to address the privacy-aware task allocation problem defined in Section \ref{sec:problem}.

\subsection{\zy{System Overview and Workflow}}

\begin{figure}[!htb]
    \centering
    \includegraphics[width=\linewidth]{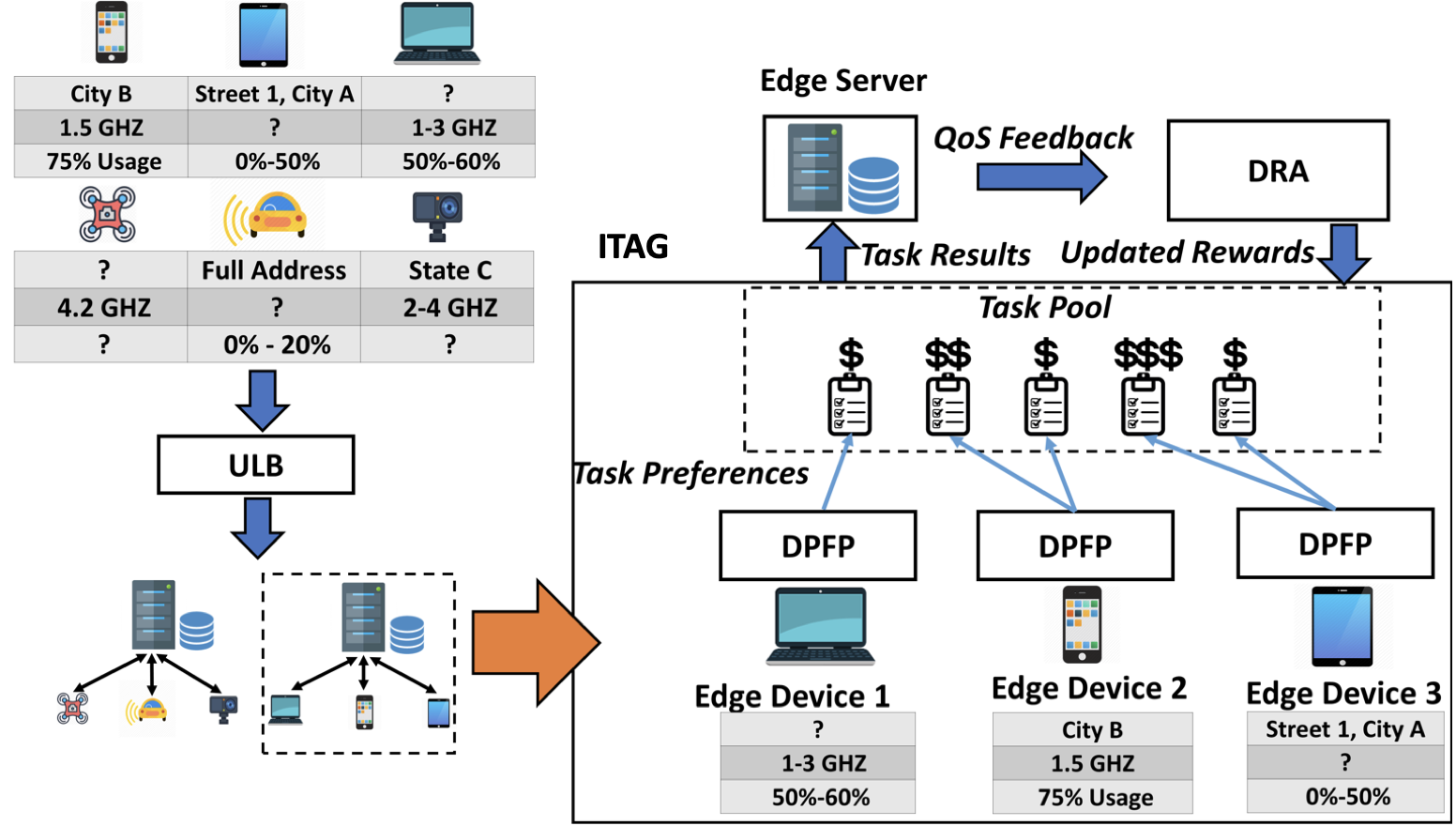}
    \caption{Overview of G-PATA}
    \label{fig:overview}
\end{figure}

An overview of G-PATA framework is given in Figure~\ref{fig:overview}.  The G-PATA framework consists of four major components: (i) an Incentive-driven  Task Allocation Game (ITAG)  that  models the cost and benefits of each task allocation strategy for the end devices; (ii) a Decentralized  Privacy-aware Fictitious Play (DPFP) scheme that aims at maximizing the individual payoffs of end devices while respecting the privacy requirements of the end users;  (iii) a Dynamic Reward Assignment (DRA) scheme that dynamically tunes the rewards to  ensure the QoS of the application; and (iv) an Uncertainty-aware Load Balancing (ULB) scheme that  efficiently assigns end devices to edge servers to make the system adaptive to the dynamics of the end devices. We present these components in detail below.


\subsection{An Incentive-driven Task Allocation Game}
The task allocation problem can be modeled as an bottom-up Incentive-driven Task Allocation Game (ITAG) where the end devices, rather than the server, decide their preferred tasks to selfishly maximize their individual payoffs.
The rationale of adopting a bottom-up game-theoretic approach is mainly three-fold: 1) \dz{the server has only a limited amount of information of the end devices to make the globally optimal task allocation decisions;} 2) the game theory approach allows the end devices to collectively achieve maximum payoffs for themselves, thus facilitating the compliance of the end users; 3) the bottom-up approach eliminates the necessity of sharing private information to the server for decision making, which provides the foundation for achieving privacy-awareness in SSEC.

\subsubsection{Game Setup}


\dz{We first describe how the ITAG is set up. Recall that in SSEC, we have a total of X end devices, \dzf{M} tasks and Y edge servers as described in Section \ref{sec:problem}. The ITAG is played with a local cluster of end devices -- an edge server and the end devices  connected to it. We discuss how the local cluster is formed in Section \ref{sec:balance}.}

\dz{An ITAG  can be described by a tuple $(S_y, E\!D_y, T\!K, A, \Psi,\pi, R)$ where $S_y \in E\!S$ is the local game host (i.e., an edge server) and $E\!D_y\in E\!D$ is a set of $J$ players (end devices that are assigned to $S_y$) $\{E_{y,j}| 1 \le j \le J, E_{y,j} \in E\!D\}$, where $1 \le J \le X$.}\footnote{We omit the index for sensing cycle (i.e. $t$) and index for edge server (i.e., $y$) for ease of notation in the rest of the section.} $T\!K$ is a set of $I$ computation tasks $\{\tau_i| 1 \le i \le I, \tau_i \in Tasks\}$ where $1 \le I \le M$.    In each sensing cycle, we define a Strategy Profile $A$ as a set of individual task allocation strategies, i.e. $A = \{a_1, a_2,...,a_J\} \in \Psi$, where $a_j$ denotes the task strategy (i.e., which tasks to execute) on the end device $E_j$. For example, if a local edge server is connected by 2 end devices and there are a total of 4 tasks. end device 1 picks task 1 and 3 while end device 2 picks task 2 and 4. Then we have $a_1= \{1,3\}$ and $a_2 = \{2, 4\}$. We use $\Psi$ to denote the computation task allocation strategy space for all players: $\Psi = \Psi_1 \times \Psi_2 \times... \times \Psi_{J}$ where $\Psi_j$ represents the strategy space of player $E_j$ (i.e., $2^I$ possible task allocation strategies for each device). $\pi$ represents a device-specific cost function vector $\pi = \{\pi_1,\pi_2,..,\pi_J\}$ for all end devices where the element $\pi_j$ represents the cost of the end device $E_j$ to execute the allocated tasks based on strategy $a_j$.   $R = \{R_1, R_2...,R_I\}$ is a set of rewards that is provided by the application where $R_i$ represents the reward for executing task $\tau_i$. \dz{We further define a payoff function for a task allocation strategy  $a_j$ as a function $g(\cdot)$:}

\begin{equation}
u_{j} =  g(R,\pi_j,a_j, a_{-j}).
\end{equation}\noindent
Here $a_{-j}$ represents the strategies for other end devices. The payoff function represents how much benefits the end device $E_j$ can gain if strategy $a_j$ is taken by $E_j$. The objective of  $E_j$ is to pick the best task allocation strategy that maximizes its own payoff function, given the strategies picked by other end devices: 
\begin{equation}\label{eq:payoff2}
\operatorname*{argmax}_{a_j}~   g(R,\pi_j,a_j, a_{-j})
\end{equation}\noindent

\subsubsection{Game Protocol}

The high-level game protocols of every ITAG at each sensing cycle is designed as follows: (i) the edge server broadcasts the tasks and the reward for each task to the end devices \dz{at the beginning of every sensing cycle;} (ii) each end device simultaneously picks a task that has the best payoff for itself; (iii) if multiple nodes choose the same task, the server performs a \emph{tie breaking process} to assign a task to a specific device. We refer to one round of this task allocation process as an ``iteration'', and assume that each end device is myopic and only picks one task at a time (singleton property) within each iteration; (iv) keep iterating until all tasks are picked. After all the tasks are assigned,  the end devices start to process the tasks they picked. 

\dz{The game protocol described above has two key properties: 1) \emph{congestion property} -  the expected reward of each task monotonically decreases as the number of end devices that pick the task increases; 2) \emph{singleton property} - each player can only pick one task at a time. The key benefit of these two properties is the guaranteed existence of Nash Equilibrium \cite{ackermann2006pure} which is crucial for the end devices to make mutually satisfactory task allocation decisions. The singleton property reduces the strategy space $O(2^I)$ to $O(I)$ for each iteration which significantly reduces the task allocation overhead. }


\subsubsection{Tie Breaking Process and Payoff}

\zy{In real-world scenario, the amount of end devices is enormous and it is very likely that multiple nodes choose the same task. The \emph{tie breaking process} described in the game protocol is a critical component in ITAG.  It allows the edge server to decide which end device to execute the task that yields the best system performance when multiple devices are competing for the same task. However, picking the best assignee is challenging because the edge server can only observe partial information from the end devices due to privacy constraints. For example, if two devices are competing for the same task, it is best to assign the task to the device with better computing power (to reduce execution delay) or closer distance (to reduce network delay). But it is likely that neither device is willing to share their computing power or location. To perform tie-breaking, we first design a mechanism to determine the \emph{quality} of each end device. We formally define the quality of an end device as:}

\begin{myDef}
\emph{\dz{Quality of end device $q_{i,j}$:}  a score metric denotes how efficient an end device $E_i$ can finish a task $\tau_j$. The efficiency is measured in terms of computing power and transmission delay. The higher computing power and lower transmission delay a device has, the higher the quality score should be. }
\end{myDef}

\dz{Note that in the above definition, the quality score also depends on which task is assigned to the device. This is because the transmission delay and computing power have to be evaluated in the context of which task it is going to execute. An end device has different computing capabilities and transmission delays for different tasks. }

\dz{The server derives a \emph{quality score $q_{j,i}$} for each device as:}
\begin{equation}
q_{j,i} =  \lambda_1 \cdot \widehat{f_{j}} \cdot  (1- \widehat{u_{j}}) - \lambda_2 \cdot \widehat{dist(E_j,S(j))}.
\end{equation}\noindent
where $\widehat{f_{j}}$ and $\widehat{u_{j}}$ denotes the estimated CPU frequency and usage for $E_j$. $\widehat{dist(E_j,S(j))}$ denotes the estimated distance from $E_j$ and the assigned edge server $S(j)$. $\lambda_1, \lambda_2$ are tunable parameters. 

Then, to perform tie breaking, the server randomly assigns the task $\tau_i$ to $E_j$ with probability 
\begin{equation} \label{eq:quality2}
\frac{q_{j,i}}{\sum\limits_{j' \in D(i)}q_{j',i} }
\end{equation}\noindent
\dz{where $D(i)$ denotes the set of devices that pick $\tau_i$. The intuition is that end devices with shorter distances to the edge server (lower latency) and higher computation power is preferred. The tie breaking strategy based on Equation \eqref{eq:quality2} allows computationally inferior end devices to refrain from taking tasks with high rewards (i.e., computationally expensive tasks). }

\dz{We now show how the server estimates the status information, (i.e., CPU frequency and distance) of end devices given the partial information shared by the end device. First, we model the partial device information as a \emph{probability density function} (PDF) over a specific \emph{uncertainty region} $UR = [LB, UB]$. This is a classical technique for modeling uncertain data \cite{ngai2006efficient}. For CPU frequency and usage information, the uncertain region is the frequency/utilization range that a user shares to the application. For example, assume an end device discloses its CPU frequency as 1-3 GHz and CPU usage as 50\% - 100\%, then the uncertainty regions are [1,3] and [50\%, 100\%], respectively. The PDF is a  distribution over the region. For example, the server can assume a PDF as a  uniform distribution which means the server can only randomly guess the actual CPU frequency and its usage in the corresponding range.  For location information, the uncertainty bound is a geographical region that a user is willing to share. For example, if an end device discloses its location as State A,  then the geographical region of State A is the uncertainty bound. The PDF can be estimated based on the prior knowledge of the region. For example, the PDF can be assumed to be the population distribution (intuitively, the end device has a higher chance to locate in an area with higher population density) over State A.}

\dzf{Based on the above model for uncertain information, the server can estimate the end device's status using one of two methods. The first method is \emph{anchor point based estimation approach} \cite{chau2006uncertain}. In this method, the edge server first samples a total of $K$ values from PDF within the uncertainty bound (referred to as anchor points) and then estimates the device status and location of end devices based on the average of the sampled K anchor points. The second method is a \emph{conservative information estimation approach}. In this approach, the edge server estimates the CPU frequency, CPU usage,  and location of end devices based on the worst-case estimate \cite{martin2007worst}. To be specific, for location, we assume the end device has the furthest distance within UR to the edge server. For CPU operating frequency and usage, we assume it has the lowest frequency and highest usage within UR. }
\dz{In this paper, we pick the conservative information estimation to run on the edge servers because 1) the anchor point based estimation may encourage lower-end end devices to hide their private information and keep themselves from being selected for the task allocation; 2) conservative information estimation encourages end devices to trade their privacy for more incentives. } \dzf{For example, if an end device A has a frequency of 1 GHz while the other devices are powerful and have an average frequency of 2.5 GHz. When the server uses anchor point based evaluation, device A will not be able to compete for tasks if it reports the exact device frequency. Instead, device A will be tempting to report a vague range such as [0, 10] GHz to trick the system to better compete for rewards. Conservative information estimation solves this problem by adopting the worst-case estimation so that device A has no incentives to report a vague range rather than the actual frequency. 
}

\dz{After introducing the ITAG game protocol and tie breaking process, we can now define the key component in the bottom-up task allocation process - the \emph{payoff function}. Let's assume the device $E_j$ picks the $i^{th}$ task in an iteration. We can define the individual \emph{payoff function} of $E_j$ as:}
\begin{equation} \label{eq:payoff}
\begin{split}
u_{j,i} &= \left\{
\begin{aligned}
& \frac{R_i  \cdot q_{j,i} }{e_{j,i} \cdot \sum\limits^{j' \in D(i)}q_{j',i} }, \hspace*{1em} \frac{c_i^j}{\Delta} + \widehat{u_{j}} \le 1\\
&0,\hspace*{1em} \frac{c_i^j}{\Delta}+ \widehat{u_{j}}>1
\end{aligned}
\right.
\end{split}
\end{equation}
\noindent
where $u_{j,i}$ is the payoff function of $E_j$ if it picks edge task $\tau_i$ in the current iteration. $e_{j,i}$ is the energy cost of the $i^{th}$ task of end device $E_j$.  Note that more comprehensive cost functions can be easily plugged in into this framework. $c_i^j$ is the  estimated WCET of $\tau_i$. $\widehat{u_{j}}$ captures the total utilization of all the tasks currently assigned to  $E_j$. We assume that earliest deadline first (EDF) scheduling algorithm is used at $E_j$. With EDF, task would miss its deadline if $\frac{c_i^j}{\Delta}+ \widehat{u_{j}} >1$. In the above payoff function, we discourage the end devices to pick tasks that they could not finish before the deadline by providing no reward for those tasks.


To ensure that each end device makes its best decision towards maximizing the above payoff, \zy{our goal is to find a Nash Equilibrium for the ITAG.  The Nash Equilibrium exists in a non-cooperative game where each player (i.e., end device in our model) has nothing to gain by only changing his/her own strategy.} In the next subsection, we describe how to find the Nash Equilibrium in ITAG in a decentralized and privacy-aware fashion.

Note that the payoff function (in Equation \eqref{eq:payoff}) of an end device depends on the aggregated \dz{quality scores of other end devices} (i.e., $\sum\limits^{j' \in D(i)}q_{j',i}$) without the necessity of understanding the exact payoffs of other end devices. This is an important property we used for designing a privacy-aware fictitious play algorithm discussed below.

\subsection{Decentralized Privacy-aware Fictitious Play} 

In this subsection, we develop a  Decentralized Privacy-aware Fictitious Play (DPFP) solution for SSEC that finds the Nash Equilibrium in an iterative fashion.  The design objective of the DPFP is 3-fold: 1) no information other than that willingly shared by the end devices is used to perform task allocation; 2) the decision privacy must be ensured such that no individual device's strategy is disclosed; 3) the algorithm can  efficiently adapt to time-critical social sensing application with constantly changing device status. 

Classical Nash Equilibrium (NE) solutions, such as best response dynamics, reinforcement learning, or Fictitious Play~\cite{bowling2002multiagent} cannot be applied to the ITAG model. This is because each end device has only partial information of other end devices due to the privacy constraints, thus cannot precisely derive other devices' best strategies to make its own best decision (similar as in an auction scenario where each bidder would not share her own valuation of an item beforehand). Moreover, these algorithms require all end devices to share their exact strategies in the game, which violates the decision privacy defined in Section \ref{sec:problem}.

Our solution develops a \emph{Joint-Strategy Best Response Dynamics}  scheme where each end device predicts the aggregated ``congestion rate" of a computation task instead of the actual individual strategy (i.e., the task being picked) of an end device. Therefore, an end device only observes the number of devices that plan to pick a specific task rather than the actual strategy of other devices. Our scheme enables an obfuscation via the grouping of the strategies from end devices, ensuring decision privacy.    

Specifically, at each iteration of ITAG, each end device first estimates a congestion rate as $\mathcal{N}_i = \widehat{\sum\limits_{j' \in D(i)} q_{j,i}}$. The congestion rate is the estimation of the overall quality score of all devices that may pick $\tau_i$ in the next sensing cycle. The congestion rate is dynamically estimated as: 
\begin{equation} \label{eq:histo}
\mathcal{N}_i = \mu \cdot \mathcal{N}_i^{prev} + (1-\mu) \cdot \sum\limits_{j' \in D(i)} q_{j,i}
\end{equation}
where $\mu \in (0,1]$ is a decay factor that controls the importance of more recent observations. $\mathcal{N}_i^{prev}$ is the estimated congestion rate in the previous sensing cycle and $\sum\limits_{j' \in D(i)} q_{j,i}$ is the actual congestion rate of $\tau_i$ in the current sensing cycle, which is synchronized to the end devices from the edge server.

Then, based on the estimation, each end device finds the best response given the estimated congestion rates. Intuitively, if a task is predicted to be too ``congested", the end device may want to avoid picking that task (even the task has a high reward), and resort to picking less congested ones. The best response is derived as: 
\begin{equation} \label{eq:bestr}
\begin{split}
 a^*_j = \left\{
\begin{aligned}
&\operatorname*{arg\,max}_{i} \frac{ R_i  \cdot q_{j,i} }{e_{j,i}\cdot \mathcal{N}_i}, with~probability~\rho\\
 & a_j^{prev} , ~with~probability~1-\rho
 \end{aligned}
\right.
 \end{split}
\end{equation}
\noindent
where $a_j^{prev}$ denotes the strategy of $E_j$ in the previous iteration. $\rho$ is an ``inertia term" that allows the end devices to be ``lazy" at certain rounds of the game (i.e., the end device do not need to update their strategy at every iteration) so the communication overhead can be significantly reduced.

\zy{The ITAG continues the above process until a convergence signal is found - i.e., the server found no end device would change its strategy in two consecutive iterations. Then ITAG concludes a Nash Equilibrium is found. We refer the proof of convergence of FP with inertia and the existence of Nash Equilibrium in \cite{swensondistributed}.}

\begin{figure*}[!th]
\centering
\begin{minipage}{0.9\textwidth} 
\includegraphics[width=\linewidth]{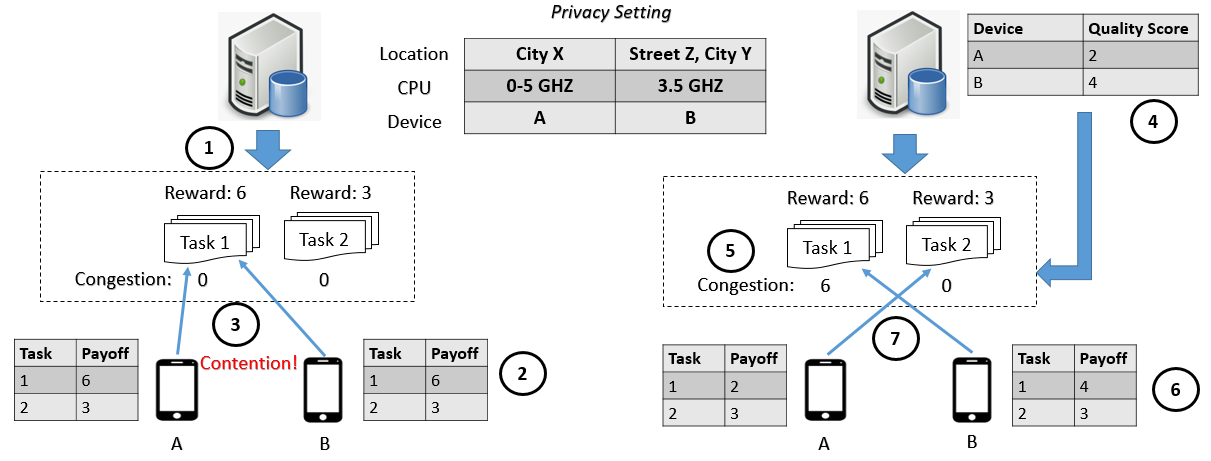}
{\small \rv{Workflow - \raisebox{.5pt}{\textcircled{\raisebox{-.9pt} {1}}} The edge server first assigns tasks and their associated rewards to the end devices. \raisebox{.5pt}{\textcircled{\raisebox{-.9pt} {2}}} The end devices calculate the payoff function of each task. \raisebox{.5pt}{\textcircled{\raisebox{-.9pt} {3}}} Both end devices believe Task 1 yields the highest payoff of 6, causing them to compete for Task 1. \raisebox{.5pt}{\textcircled{\raisebox{-.9pt} {4}}} The edge server observes the devices' choices, calculates a quality score based on the uncertain information (due to privacy protection) of the device. Device B is assigned a higher score as it provides more information to determine its CPU power and distance.  \raisebox{.5pt}{\textcircled{\raisebox{-.9pt} {5}}} The edge server updates a congestion rate for each task, representing the total quality scores of the devices that pick the  task. \raisebox{.5pt}{\textcircled{\raisebox{-.9pt} {6}}} Each end device updates its payoffs after observing the congestion rate. Device A finds that its quality score is relatively low as compared to the congestion rate (overall quality score) of Task 1, thus discounting the payoff for Task 1. \raisebox{.5pt}{\textcircled{\raisebox{-.9pt} {7}}} After the payoff being recalculated, Device A gives up on Task 1 and picks up Task 2 instead. The algorithm converges with no contention.}
\par}
\end{minipage}
\caption{Illustrative Example of ITAG}
    \label{fig:toy}
\end{figure*}

One key limitation of FP-based solutions is the slow convergence rate, which would result in significant performance degradation, thus seriously impact the QoS of real-time systems \cite{danielrtas}.  To further improve the efficiency of the DPFP algorithm, we set an upper bound on the number of iterations. We refer to this modification as \emph{P-chance Negotiation} where each end device has a maximum number of P times to regret and re-adjust their strategy. Note that when P=1, the scheme degenerates to a \emph{sealed bidding game} \cite{varian2014vcg} where each end device can only negotiate once. 
 We summarize the DPFP algorithm in Algorithm \ref {alg:dfp}. We also illustrate a toy run of two devices competing for 2 tasks in Figure \ref{fig:toy}.

\begin{algorithm} 
    \footnotesize
    \caption{Decentralized Privacy-aware Fictitious Play}
    \label{alg:dfp} 
    \begin{flushleft}
    \hspace*{\algorithmicindent} \textbf{Input}: $T\!K$, $R$, $I$, $J$, $\Delta$, $P$\\
    \hspace*{\algorithmicindent} \textbf{Output}: task allocation Strategy for $E_j$
    \end{flushleft}
    \begin{algorithmic} [1]
        \State    Initialize:  $taskCount$ $\leftarrow new Array [J][I]$, $A$ $\leftarrow new Array [J]$,          $converge \leftarrow$ $False$
        \State Predict congestion rate of all tasks based on Equation \eqref{eq:histo}  \Comment{Congestion Rate Estimation}
                \While {$converge \neq True$~or~$iterCount \le P$}
        \State $iterCount \leftarrow iterCount+1$        
        \State find $h \leftarrow$ optimal strategy for $E_j$  based on Equation
        \eqref{eq:bestr}  \Comment{Best Response}
        \State send $h$ to server, receive congestion rate of all tasks, convergence signal $sig$
        \State $converge \leftarrow sig$
        \State update congestion rate prediction based on Equation 
        \eqref{eq:histo}
        \EndWhile              
        \State Return $A[j]$  
    \end{algorithmic}
\end{algorithm}

\subsection{Dynamic Reward Assignment}

The ITAG and DPFP approaches discussed above ensure that the end devices achieve maximum payoff through non-cooperative gameplay. As shown in the payoff functions (Equation \eqref{eq:payoff}), the definition of the rewards plays a critical role with respect to whether a computation task is picked by end devices for execution. Due to the dynamic compliance issues, (i.e., the fact that the cost of end devices and their willingness to execute tasks can change over time), 
static reward functions are generally not optimal for the server to satisfy QoS requirements \cite{danielrtas}. 
We introduce a Dynamic Reward Assignment (DRA) scheme for the server to dynamically update task rewards in order to meet the QoS objective of the application.  Below, we first present the general reward function and then discuss how it is dynamically adjusted.

The DRA performs a 2-step control process: 1) DRA dynamically decides reward distribution among tasks; 2) DRA dynamically adjusts the overall budget for the next sensing cycle based on the performance feedback.

Given the budget, the server calculates the reward of a task $\tau_i$ based on the computational complexity and transmission complexity to be processed:

\begin{equation} \label{eq:reward}
R_i =  \frac{\alpha_1}{\alpha_1 + \alpha_2} \cdot O_i^{c}  + \frac{\alpha_2}{\alpha_1 + \alpha_2}\cdot O_i^{t}  
\end{equation}
where $O_i^{c} $ and $O_i^{t} $ denotes the expected computation and communication complexity respectively. $\alpha_1$ and $\alpha_2$ are the weighting factors. The task rewards are normalized so that the reward of each individual task is bounded. The design of the task reward function is intuitive: higher computational complexity results in higher energy consumption on the end device, thus should be compensated for higher rewards; same intuition for transmission complexity.

After the tasks are executed, the server can calculate the control signals based on which tasks missed the deadlines and adjust the weights accordingly using a exponential weight update rule \cite{cesa2006prediction}:

\begin{equation}\label{eq:weight}
\alpha_k^{new} = \alpha_k^{old}*e^{-\eta \mathcal{L}_k} , \hspace*{1em} k = 1~or~2.
\end{equation}\noindent
where $\eta$ is a learning parameter, $\mathcal{L}_k$ is a loss function for $\alpha_k$:
\begin{equation} 
\label{eq:lambda}
\begin{split}
&\mathcal{L}_1 = \frac{\sum_{i=1}^{I} O^{t}_i  * \delta_i}{\sum_{i=1}^{I} \delta_i} - \frac{\sum_{i=1}^{I} O^t_i * (1- \delta_i)}{\sum_{i=1}^{I} (1-\delta_i)}, \sum_{i=1}^{I} \delta_i\neq 0~or~I \\
&\mathcal{L}_2 = \frac{\sum_{i=1}^{I} O^{c}_i  * \delta_i}{\sum_{i=1}^{I} \delta_i} - \frac{\sum_{i=1}^{I} O^c_i * (1- \delta_i)}{\sum_{i=1}^{I} (1-\delta_i)}, \sum_{i=1}^{I} \delta_i\neq 0~or~I \\
\end{split}
\vspace{-0.1in}
\end{equation}\noindent
where  $\delta_i$ signifies whether task $\tau_i$ misses the deadline or not (see Section \ref{sec:problem}). In the above equation, we use the average transmission complexity $O^t_i$ of the tasks that meet the deadlines (i.e., $\frac{\sum_{i=1}^{I} O^t_i  * (1- \delta_i)}{\sum_{i=1}^{I} (1-\delta_z)}$) as a set point and compare it with the average transmission overhead of the tasks that missed the deadlines ($\frac{\sum_{i=1}^{I} O^t_z * \delta_i}{\sum_{i=1}^{I} \delta_i}$). 
The same idea applies for the computation factor (using computation overhead $O^c_i$). 
The parameters  $\alpha_1$ and $\alpha_2$ are tuned to minimize the loss functions.

The DRA then adjust the overall budget via a simple feedback control:
\begin{equation} \label{eq:budget}
\begin{split}
budget_{new} &= \left\{
\begin{aligned}
& budget_{old} \cdot \beta, \hspace*{1em} \sum_{i=1}^{I} \delta_i \ge thres_{high} \\
& budget_{old},\hspace*{1em} \sum_{i=1}^{I} \delta_i < thres_{high}\\
\end{aligned}
\right.
\end{split}
\end{equation}

In the rare case where all jobs miss their deadlines (i.e., $\sum_{z=1}^{Z} \delta_z= Z$), we resort to admission control and decrease the number of assigned jobs. Due to the page limit, we omit the detailed discussions.

\subsection{Efficient Load Balancing with Uncertainty-awareness}\label{sec:balance}

We further propose a load-balancing design for G-PATA where the end devices are assigned to multiple edge servers. We refer to the edge server and the end devices connected to it as a ``local group". The assignment of an end device to an edge server is referred to as an ``edge-server" assignment. By deploying more edge servers, the end devices in SSEC can be split into multiple local groups  with manageable sizes.

\dzf{The above load balancing design is particularly necessary for the G-PATA due to the scalability requirement. In particular, we found the complexity of the ITAG model in G-PATA largely depends on the number of end devices that participate in the task bidding process. Therefore, a potential scalability issue rises: as the number of participating devices increases, the overhead (including both execution time and the communication overhead between the end devices and the edge server during the negotiation process in DPFP) of the task allocation process becomes excessive. To allow the G-PATA framework to be scalable, we design a load balancing module where the ``load" refers to the overhead incurred by the task allocation process (i.e., ITAG model). The load balancing module can judiciously assign end devices to the corresponding edge servers \dzf{to split the load of executing the task allocation model (i.e., ITAG).} Each edge server plays the ITAG with the end devices \dzf{that are assigned to it}. }

Note that a naive load balancing design such as assigning end devices to the nearest edge servers will not work as  users and servers are usually distributed unevenly within a network, thus resulting in severe unbalanced load. Additionally, the load balancing should take into account the privacy requirement of the SSEC system. In particular, the location of the end devices are \emph{uncertain}, thus it is hard to precisely identify the nearest edge server for an end device. We develop a simple and effective Uncertainty-aware Load Balancing (ULB) module that can effectively balance the load of end devices without violating the privacy constraints of  the end devices. For end device $E_x$, the goal is to find its assigned edge server $S_{y,x}$ that satisfies:

\begin{equation} \label{eq:ulb1}
 \operatorname*{arg\,min}_{y} \sum_{x=1}^{N} ExpD(E_x,S_{y,x}) = \sum_{x=1}^{N} (\int f_x(r)d(E_x,S_{y,x})dr) 
\end{equation}\noindent
where $ExpD$ is the expected distance function based on a distance metric $d(\cdot)$. $f_x(r)$ denotes the probability density at $r \in UR$. In this work, we define the distance metric $d(\cdot)$ as:

\begin{equation} \label{eq:group}
d(E_x,S_{y,x}) = \frac{\widehat{dist(E_x,S_{y,x})}} {\widehat{f_{j}} \cdot  (1- \widehat{u_{j}})}
\end{equation}
\noindent

Note that finding the optimal edge-server assignment by solving Equation (\ref{eq:ulb1}) is an NP-hard problem \cite{tindell1992allocating}. To ensure the efficiency of the ULB module, we devise a greedy algorithm. In particular, given the distance metric, each edge server picks an end device with the minimum distance in a round robin fashion. The intuition of the above distance metric is that each edge server greedily picks the end devices that have the least estimated delay and highest computing power. ULB adopts the round-robin strategy to ensure the number of end devices within each local group is similar so that the allocation processes (i.e., running the DPFP algorithm) have similar expected execution time. For each edge server, it needs to evaluate all the end devices and identify the best fit through sorting based on Equation \eqref{eq:ulb1}. To ensure the server will not pick the same device, each server deletes an end device in the sorted list once the device has been picked by another server. The complexity is O(XlogX) for sorting and O(X) for deleting the device selected by other devices. The overall computational complexity of ULB module is O(XlogX + XY) where X is the number of end devices and Y is the number of edge servers (see Section \ref{sec:problem}).

\subsection{Privacy Analysis of G-PATA}\label{sec:privacy}
\rv{Finally, we discuss the privacy property of G-PATA. The G-PATA strictly protects user-defined privacy by adopting an access control mechanism where  only the data approved by the user is disclosed to the server. This prevents unauthorized information from being disclosed in the first place (e.g., if a user chooses to reveal only the  ``city" information, then the raw location data is pre-processed so that all entities in the system can only know the location of the device at the city level ). The willingly disclosed information is then coded as an uncertainty region (e.g., the whole region of the city or the range of frequency of the CPU) via a cloaking method. Therefore, from the perspective of the servers as well as other end devices, only the uncertainty region is disclosed and they cannot possibly know the exact value within the uncertainty region  \cite{hoh2007preserving,ardagna2008privacy}.} 

		\section{Implementation}\label{sec:iml}

This section presents  the hardware setup used in our experiments and our implementation of the G-PATA framework.

 \vspace{-0.05in}
\subsection{Hardware Setup}

\begin{figure}[!htb]
	\centering
	\includegraphics[width=7cm]{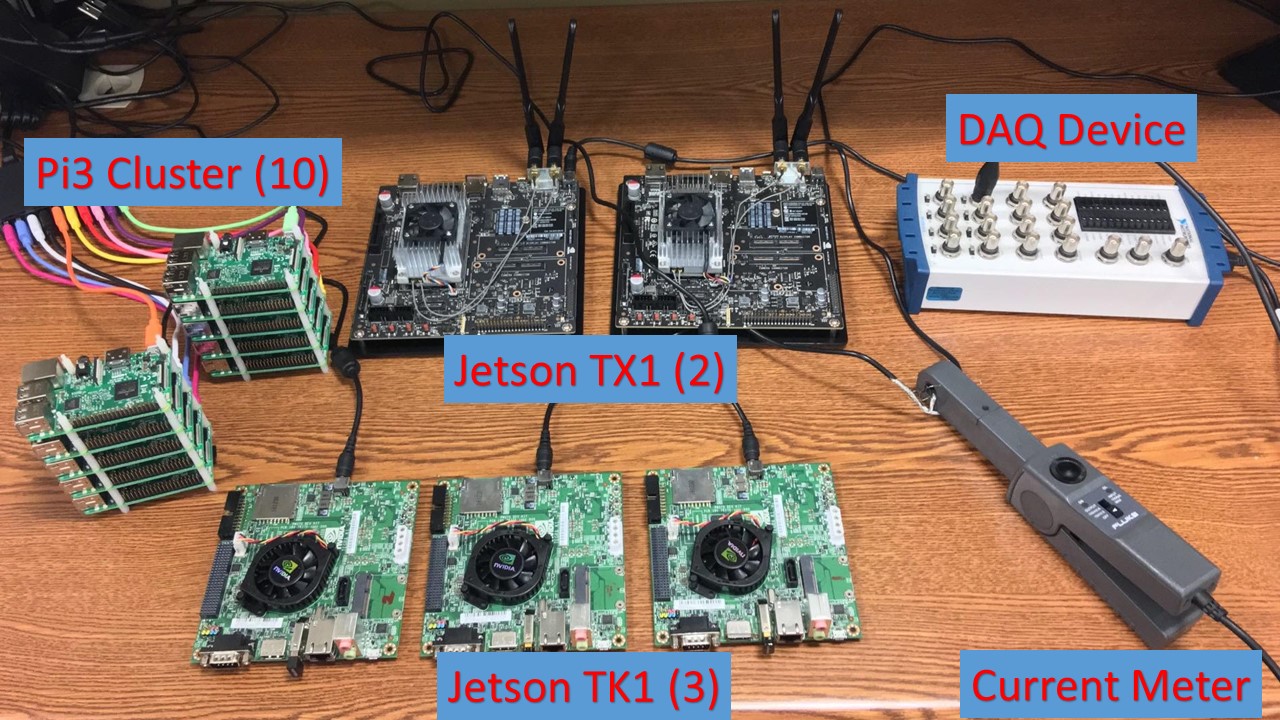}
	\vspace{-0.05in}
	\caption{SSEC Platform}
	\label{fig:edge}
	\vspace{-0.2in}
\end{figure}

 We set up a VM cluster as the remote application server on Amazon AWS by using EC2. It contains four c4.xlarge EC2 instances with each instance having four cores and 7.5 GB RAM.   We use three PC workstations with Intel E5-2600 V4 processor and 16GB of DDR4 memory as the local edge servers.  The edge servers coordinate end devices for the DPFP algorithm. Figure \ref{fig:edge} shows the hardware platform for the end devices. The platform consists of 15 end devices: 2 Jetson TX1 and 3 Jetson TK1 boards from Nvidia, and 10 Raspberry Pi3 Model B boards. \rv{These end devices have heterogeneous energy profiles and computation capabilities. The Jetson boards are commonly used in portable computers, UAVs, and autonomous vehicles and represent high-end end devices \cite{danielrtas}. On the other hand, the Pi3 boards represent lower-end devices that are very commonly used as IoT testbeds with limited computing power \cite{abbasinezhad2019novel}. All devices and the edge server are connected via a wireless router.}

\subsection{Energy Measurement}
Energy data are needed in our system. We use FLUKE AC/DC current clamps to monitor real-time current signal of each end device and capture the current values using a National Instruments USB-6216 Data Acquisition (DAQ) system. We then multiply the current values with the default voltage (12 V for TK1, 19 V for TX1 and 5 V for Pi3) to obtain the power consumption by each end device.

\subsection{System Modules and Protocol} 
We further describe the software implementation of the proposed G-PATA scheme below. 

\dzf{\textbf{Remote Server Module:} The remote server module runs on the remote application server. It has a main program that provides a global interface to the developer. In particular, it allows the user to set the overall budget, the system parameters, and monitors the status of each edge server.}

\dznew{\textbf{Edge Server Module:} The edge server module runs on each edge server. It consists of a \emph{server main program}, a \emph{load balancing program} and a \emph{DPFP server program}. The server main program performs the following tasks: i) periodically defines a set of  tasks where each task specifies what data to collect and which algorithm to use for data processing; ii) runs DRA module to adjust rewards for each task; iii) sends task information to end devices via TCP sockets; iv) receive outputs from the end devices and aggregate the results.} 
\dzf{The load balancing program performs the ULB algorithm to dynamically decide which edge server an end device should be assigned to at each sensing cycle.}  \dz{The DPFP server program coordinates the end devices in the negotiation phase of DPFP by informing the end devices of the current congestion rate of each task.}

\dznew{\textbf{Edge Modules:} On each end device, we develop three modules under the G-PATA framework: the \emph{edge main program}, the \emph{DPFP client program}, and the \emph{edge task executable}. The edge main program performs the following tasks: i) receive periodic edge task information from the server; ii) run the DPFP client program to select the  tasks it picks to execute; iii)  execute edge task executable and send results back to the server. \dzf{The DPFP client program calculates the payoff function for each end device and picks the tasks that give the highest payoff through the DPFP algorithm. The edge task executable is a standalone executable file that executes the task assigned to the end device  and generates the results.}}

	\section{Evaluation}
\label{sec:eval}

In this section, we evaluate the G-PATA scheme using the SSEC system described in Section \ref{sec:iml}. We first discuss the baselines for comparison and then present the evaluation results using two real-world social sensing applications: \emph{Collaborative Traffic Monitoring} and \emph{Abnormal Event Detection}.
\begin{table*}[t!]
\centering
\caption{Customized Privacy Levels}
\label{tab:plevel}
\begin{tabular}{|l|l|l|l|l|}
\hline
Privacy Levels &0 & 1   & 2  & 3   \\ \hline
Location       & Exact Location  & Disclose Street & Disclose City  & Hidden \\ \hline
CPU Frequency      &Exact CPU Frequency & Intervals (resolution of 0.5)   & Intervals (resolution of 2)  & Hidden \\ \hline
CPU Usage &Exact CPU Usage & Intervals (resolution of 20\%)   & Intervals (resolution of 50\%)  & Hidden\\ \hline
\end{tabular}
        \vspace{-0.1in}

\end{table*}
\subsection{Baselines}
We choose the following representative task allocation schemes from recent literature as baselines.

 \textbf{Bottom-Up Game-theoretic Task Allocation  (BGTA)}: A recent game-theoretic edge computing task allocation scheme for non-cooperative end devices. It uses a simple distributed Fictitious Play algorithm for Nash Equilibrium solution ~\cite{danielrtas}.
 
\textbf{Congestion (COG)}: A decentralized edge computing task allocation scheme using congestion game theory where the reward of a task is monotonically decreasing as more end devices claiming that task ~\cite{liu2017decentralized}.

\textbf{Greedy-Max Reward (GMXR)}: A greedy task allocation scheme where an end device greedily picks the tasks with the highest reward.

 \textbf{Top-Down Allocation (TDA)}: A top-down task allocation scheme using MILP with the objective to minimize deadline miss rate~\cite{zhu2012optimization}. The solver is running on the server. The server uses worst-case estimates of the device status based on the private information shared by end devices.

\dzf{In the above baselines, BGTA, GMXR, and COG schemes are similar game-theoretic frameworks that are completely decentralized in the sense that the servers (both remote application server and local edge servers) have no access to any of the devices' information, therefore the private information of end devices is strictly preserved. In contrast, the G-PATA scheme only protects the private information that the users are unwilling to share. Also, these decentralized algorithms require the sharing of each device's task assignment which violates the decision privacy define in Section \ref{sec:problem}. For comparison purpose, we intentionally allow such violation for two reasons: 1) we could not find relevant baselines that explicitly address such decision privacy problem and without sharing task assignment, the above baselines will not be able to run properly; 2) by comparing with these baselines, we can study whether the proposed scheme's performance will be affected by the introduction of decision privacy. }



\subsection{Evaluation Setup}
\zy{We further describe the experiment setup details.}
\zy{\subsubsection{Location and Communication Setup}
To emulate the locations of the end devices,  we use the location distribution from a Foursquare dataset collected in \cite{yang2015participatory}. From the dataset, we randomly chose three cities to emulate the sensing area of interest, where each city has a local edge server deployed. For each city, we randomly pick 5 streets, and 5 locations (i.e., point of interest (POI)) for each street. In the following experiments, we randomly assign end devices to the selected locations. Given the distance of the devices and servers, as well as the topology of the network of the SSEC system, we emulate communication delay using a NS3-based edge computing network emulator \cite{vance2018simulating}. }

\subsubsection{Privacy Level Setting}
Our work is based on the assumption that the end users are allowed to set their own privacy preferences. \dz{To empirically evaluate how well the proposed scheme performs under different privacy settings, we defined 4 levels of privacy requirements. Each privacy level is associated with different resolutions of privacy from the fine-grained resolution (i.e., less privacy protection) such as exact location and CPU frequency  to the coarse-grained resolution (i.e., more privacy protection) such as a range of possible CPU frequencies and locations. The privacy level settings are summarized in Table~\ref{tab:plevel}.} For example, if the CPU frequency is set as a privacy level of 1,  then we define a set of ranges (0-0.5], (0.5-1],..., (4.5-5]. The user only needs to disclose the corresponding range of its actual CPU frequency. Similarly, an end device discloses the CPU usage as one of the ranges (0\%, 20\%], (20\%, 40\%],..,(80\%, 100\%]  under the same privacy setting. \dz{For the location, the user can either choose to disclose the exact location or only disclose the street or city of the current location for obfuscation.}

\subsection{Case Study 1: Collaborative Traffic Monitoring}

\dz{In this application,  a set of drivers are tasked to use their dash cameras or smartphones to take videos of the traffic in front of their vehicles and extract relevant image features. These features can be then sent to nearby edge servers (e.g., Road-Side-Units) to further calculate the congestion rate of the road. In our experiment, we collected the video data using dash cameras from two vehicles. The data contains a total of 15 video files.  We divided the application into 100 sensing cycles and each sensing cycle processes video clips of 6 seconds. To emulate the various data volumes of each raw video data collected, we randomly sample each video clip between 1 to 15 image frames per second. A collaborative traffic monitoring computation task is defined as follows.}

\textbf{Computation Tasks for Collaborative Traffic Monitoring:}  performs feature extraction (optical flow and HOG) to the original raw image data.  The extracted feature is further processed by the edge server to infer the local traffic condition.

\dzf{The 15 video files are randomly assigned to the 15 end devices to emulate the traffic video data collected from them.} \dz{We repeat the experiments 100 times to generate the results discussed below.}

\begin{figure*}[tb]
    \centering
    \begin{minipage}{.32\textwidth}
        \centering
        \includegraphics[width=0.9\linewidth]{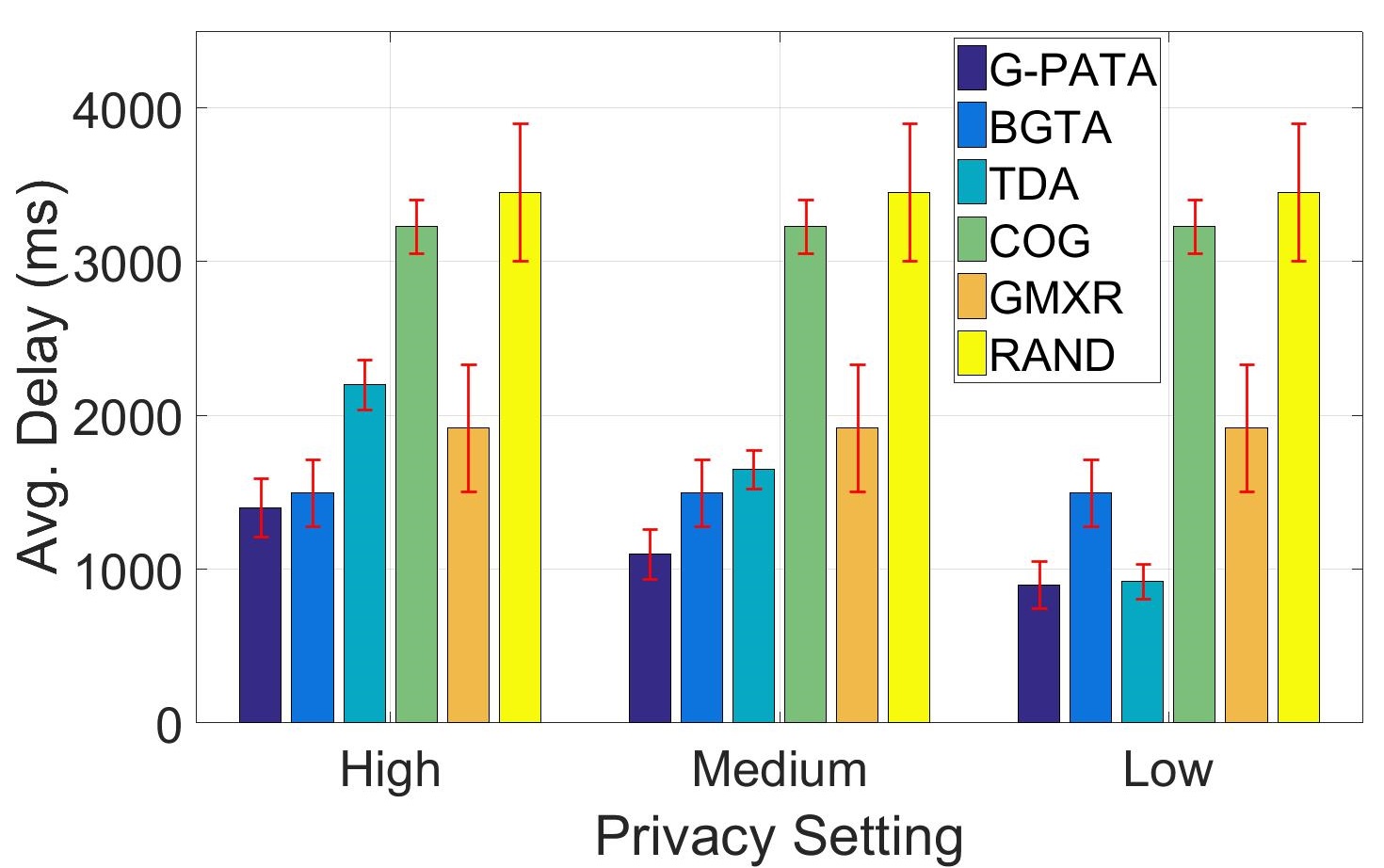}
        \captionof{figure}{Privacy Settings vs.  E2E Delay}
        \label{fig:privacytf}
    \end{minipage}%
    \begin{minipage}{.32\textwidth}
        \centering
        \includegraphics[width=0.9\linewidth, height = 0.57\linewidth ]{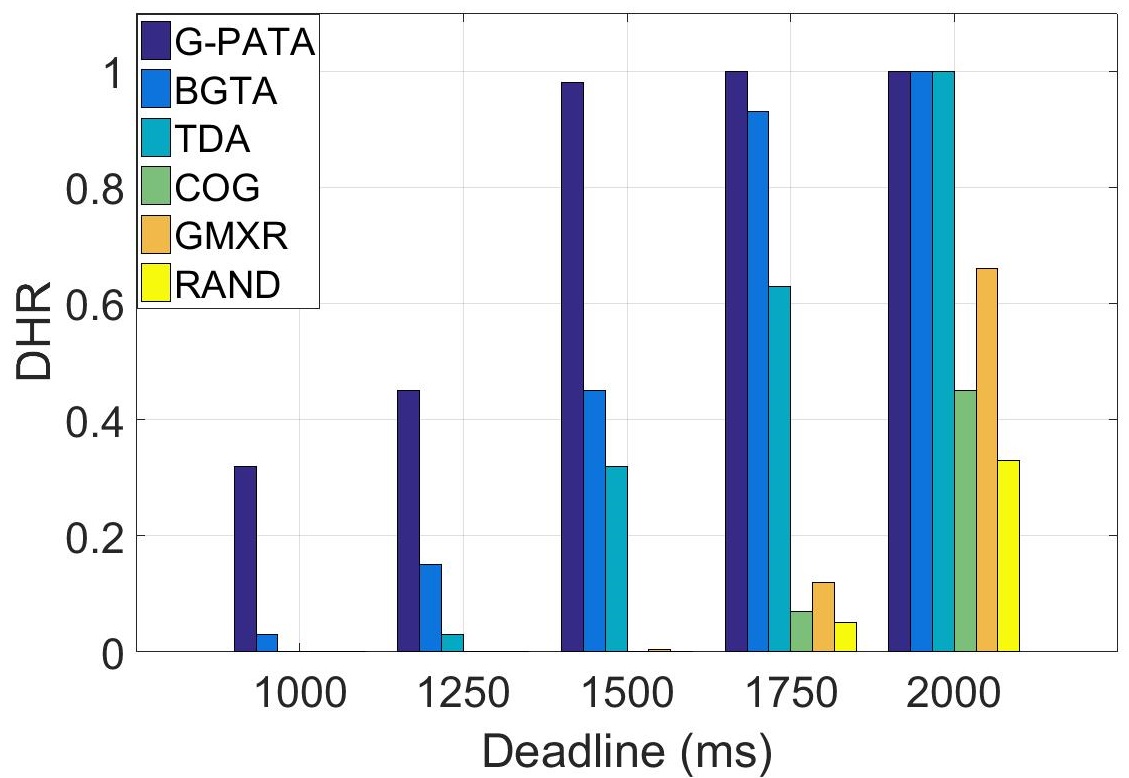} 
        \captionof{figure}{Deadlines vs. DHR}
        \label{fig:dhrtf}
    \end{minipage}
        \begin{minipage}{.32\textwidth}
        \centering
        \includegraphics[width=0.9\linewidth]{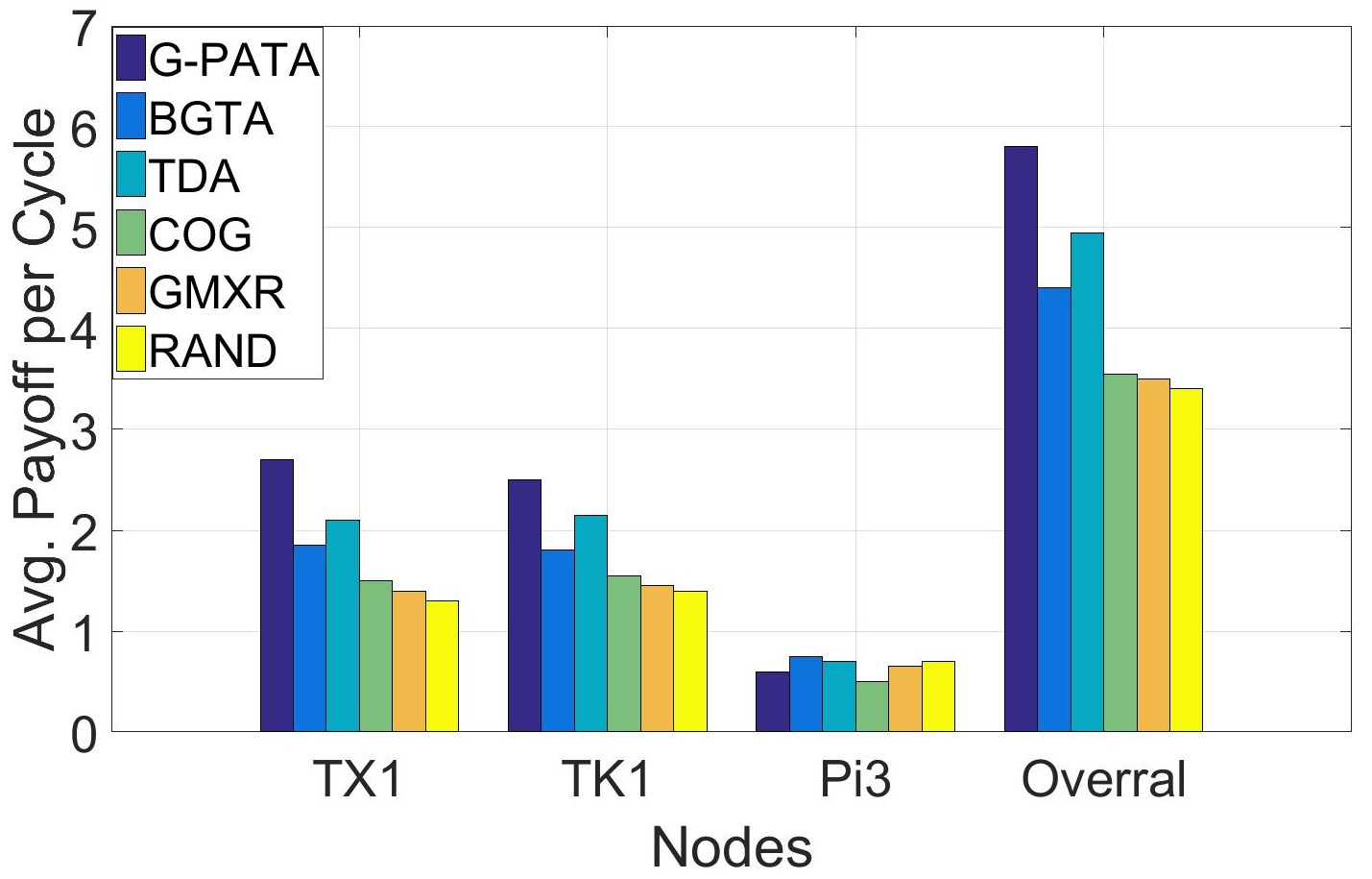}
        \captionof{figure}{Payoffs}
        \label{fig:payoff2}
    \end{minipage}
\end{figure*}

\begin{figure*}[tb]
    \centering
    \begin{minipage}{.32\textwidth}
        \centering
        \includegraphics[width=0.9\linewidth]{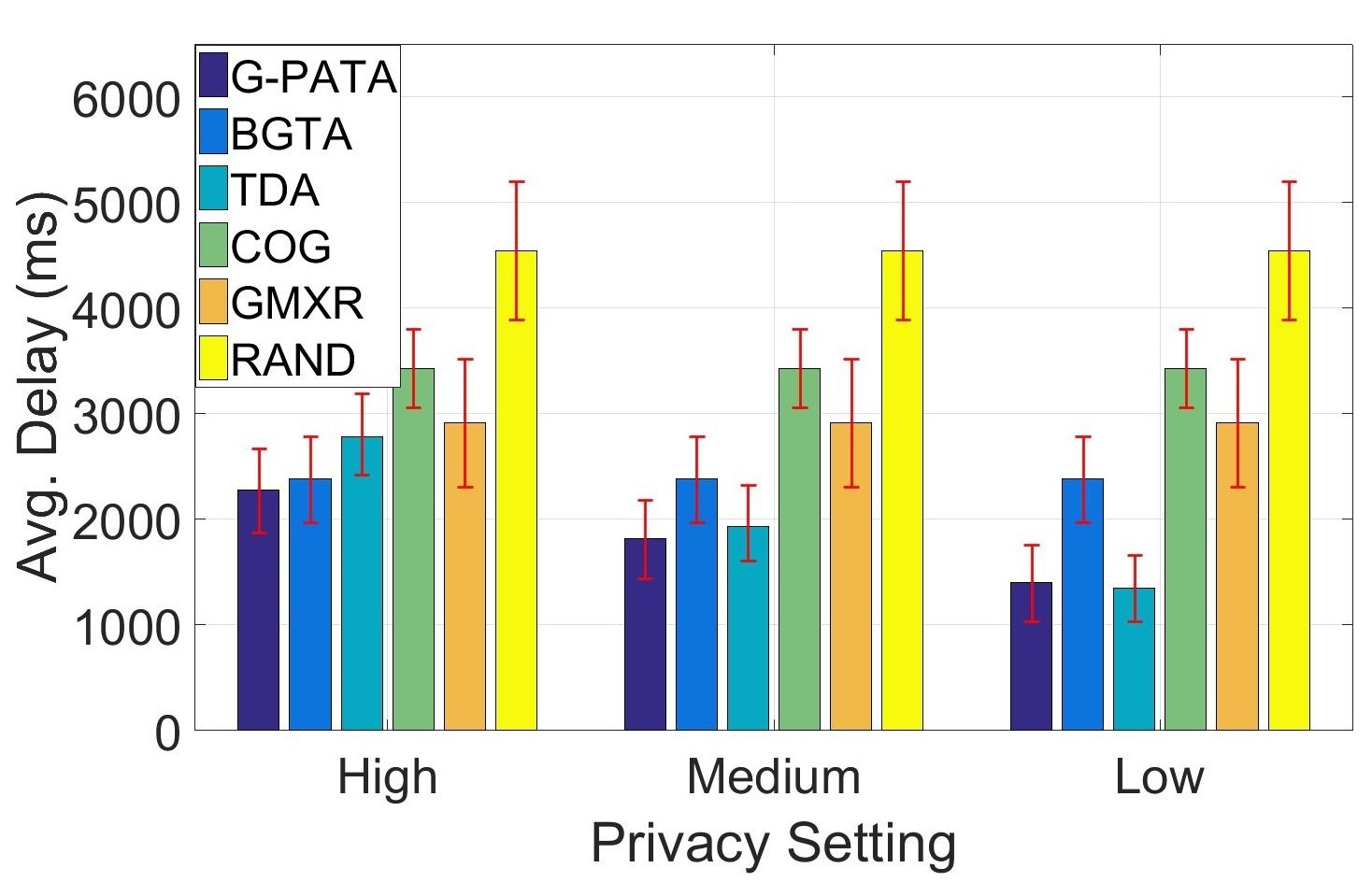}
        \captionof{figure}{Privacy Settings vs.  E2E Delay}
        \label{fig:abdelay}
    \end{minipage}%
    \begin{minipage}{.32\textwidth}
        \centering
        \includegraphics[width=0.9\linewidth, height = 0.57\linewidth ]{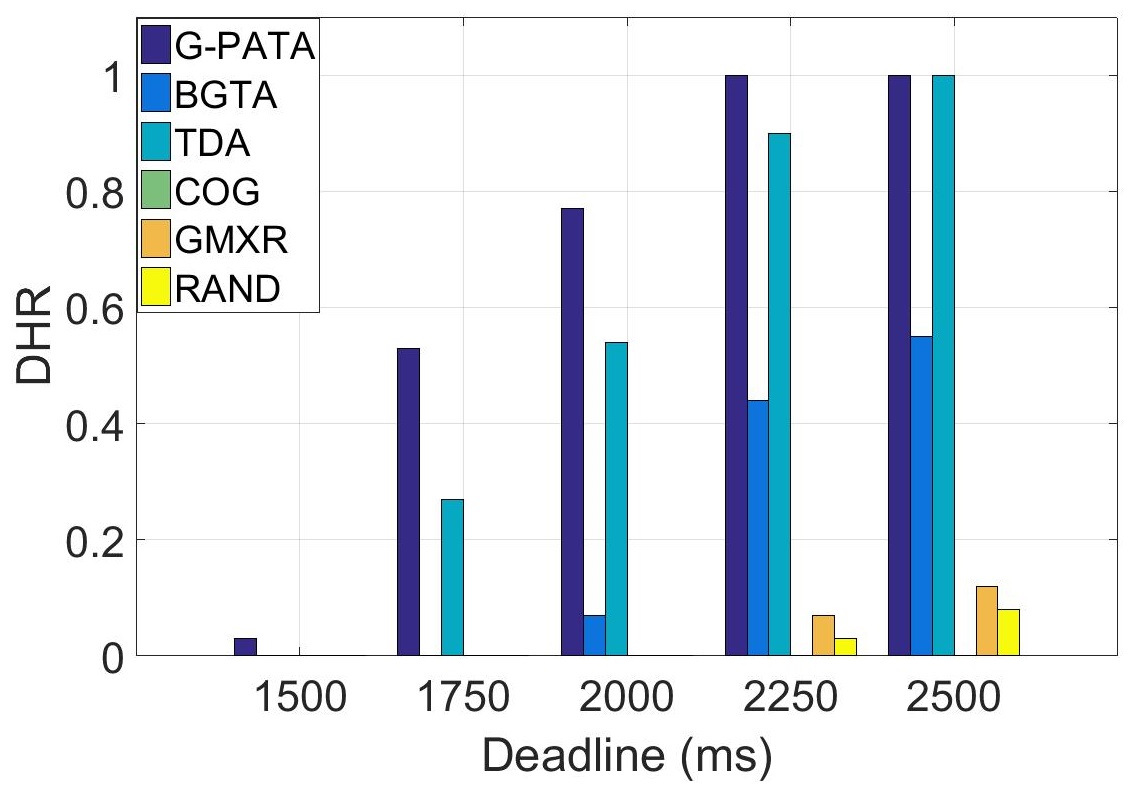} 
        \captionof{figure}{Deadlines vs. DHR}
        \label{fig:dhrab}
    \end{minipage}
        \begin{minipage}{.32\textwidth}
        \centering
        \includegraphics[width=0.9\linewidth]{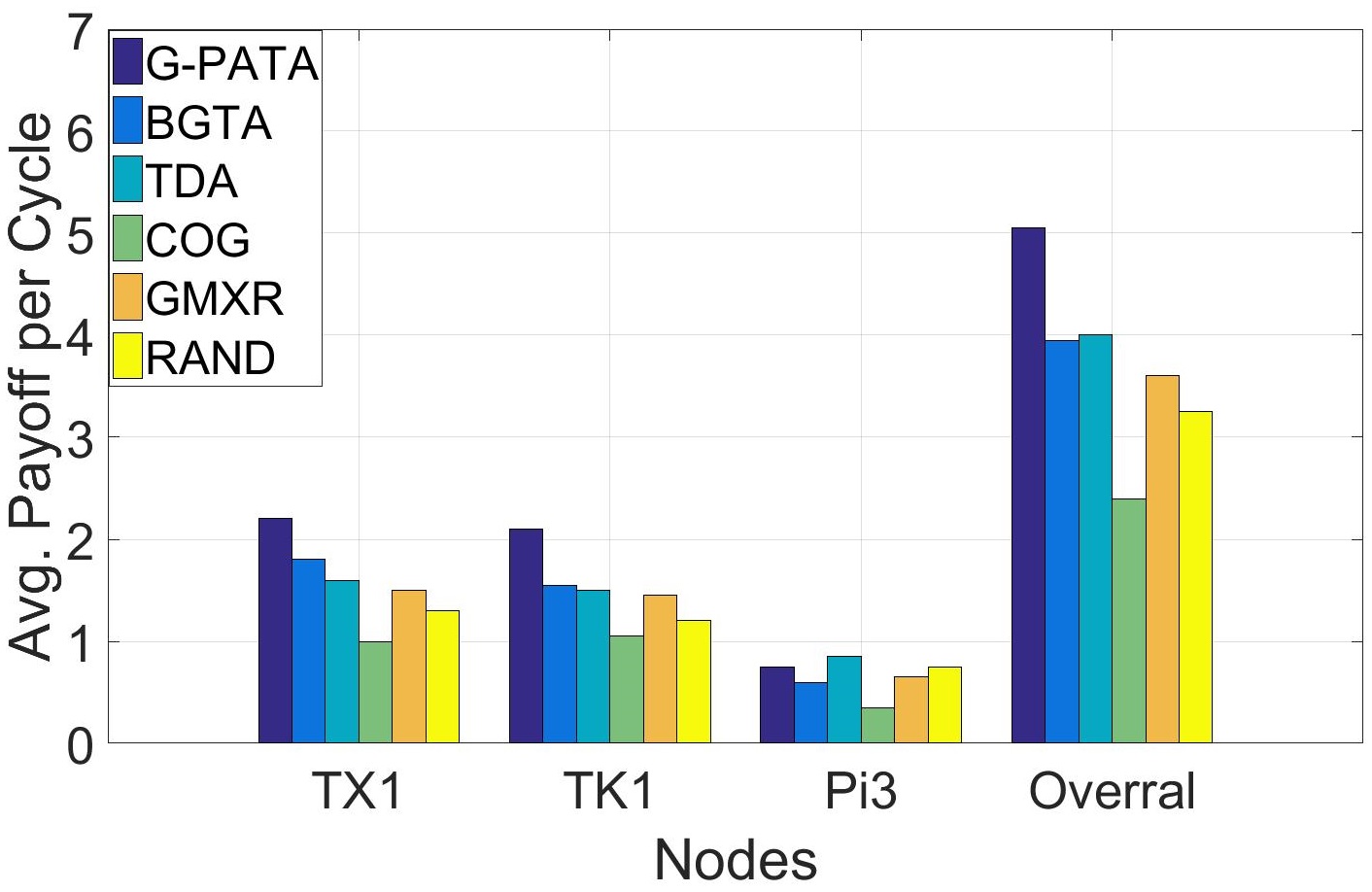}
        \captionof{figure}{Payoffs}
        \label{fig:payoffab}
    \end{minipage}
        \vspace{-0.2in}
\end{figure*}

\subsubsection{Quality of Service (Application (Server) Side)}
In the first set of experiments, we focus on how the objective is achieved from the application side. In particular, we evaluate the end-to-end (E2E) delay and deadline hit rate (DHR) of all the compared task allocation schemes. These two metrics are commonly used for evaluating the responsiveness of delay-sensitive applications \cite{danielsec}.  Figure~\ref{fig:privacytf} summarizes the E2E delays of all the schemes. We show the average delays and the 90\% confidence bounds under three scenarios of privacy settings as shown below.

\begin{itemize}
    \item \emph{High}: all end devices hide all information (i.e., choosing a privacy level of 3). 
    \item \emph{Medium}: at each sensing cycle, each end devices randomly pick its privacy level in the range of [0,3].
    \item \emph{Low}: all end devices disclose all information (i.e., choosing a privacy level of 0). 
\end{itemize}

We have the following observations: 1) our G-PATA scheme has the least E2E delay and tightest confidence bounds compared to the baselines at medium and high privacy settings; 2) we observe that G-PATA did not outperform but approaches the performance of TDA when the edge servers have full information about the end devices (i.e., under low privacy setting). This is because TDA is a top-down task allocation scheme that assumes the server has full control of the end device.  This is not practical in the edge computing paradigm and unfair to G-PATA and other baselines that assume \dz{end devices will not share all their device information with the server due to privacy concerns; } 3) the E2E performance of all schemes degrades when the privacy setting is high due to the increasing uncertainty of the end device status.

The results demonstrate the effectiveness of G-PATA for meeting real-time QoS requirements of the application while constrained by users' unique privacy requirements. The performance gain of the G-PATA is achieved by explicitly modeling the dynamic status of the end devices (e.g., computation capability and energy profile) and allocating tasks according to the current device status. We also observe that G-PATA significantly outperforms the BGTA scheme because i) G-PATA develops a privacy-aware negotiation scheme that can leverage partial information willingly disclosed by the end devices to perform better task allocation, while BGTA ignores the privacy settings and assumes the edge servers have zero knowledge of the end devices; ii) the  task allocation overhead is much less than BGTA due to the introduction of P-Chance negotiation and inertia term in the DPFP algorithm, which significantly contributes to the superior E2E delay.

\dz{Note that the privacy setting does not affect the bottom-up baselines including BGTA, COG, and GMXR, due to the fact that they completely hide the private information from the server and does not consider the customized privacy requirement from the users.} \dzf{Therefore, in the subsequent experiments, we set ``medium" as the default privacy setting for all schemes where all end devices randomly pick their privacy levels.}

The DHR is defined as the ratio of tasks that are completed within the deadline. The results are shown in Figure~\ref{fig:dhrtf}. Here we use all 15 end devices and gradually increase the deadline constraints. We fix the privacy setting as a medium. We observe that G-PATA has significantly higher DHRs than all the baselines, and is the first one that reaches 100\% DHR as the deadline increases. We attribute such performance gain to our deadline-driven dynamic incentive adjustment that guides the end devices' decisions towards minimizing deadline miss rate by providing dynamic incentives.

\subsubsection{Edge Payoffs (end device Side)}
In the second set of experiments, we focus on how the objectives of end devices are achieved. In particular, we study the payoff of end devices. Figure~\ref{fig:payoff2} shows the results of the payoff gained by the end devices.  We observe that G-PATA has the highest payoff compared to all the baselines. This is because G-PATA can most efficiently finish the jobs by pushing the computation to the edge, thus avoiding the communication latency of sending tasks to the server. This results in the least penalty for the rewards gained. Also, G-PATA allows end devices to maximize their rewards via achieving the Nash Equilibrium. The negotiation process of DPFP allows G-PATA to significantly outperform other game-theoretic schemes (i.e., BGTA and COG) by allowing end devices to collectively work out their best payoffs given their privacy constraints.

\subsection{Case Study 2: Abnormal Event Detection}

The second real-world case study is ~\emph{Abnormal Event Detection (AED)} in social sensing where the goal is to generate alerts of abnormal events from video data contributed by camera-enabled sensors (mobile phones, drones, etc.).  For example, in a mobile social sensing project, the participants are tasked to take videos/images of a location assigned by the application to help identify abnormal events such as trespassing, sudden movements, and appearance of unusual objects. Upon detection of the abnormal events, the application provides alerts to its subscribed users or the general public.

We use the UCSD Anomaly Detection Dataset~\cite{mahadevan2010anomaly} which consists of 98 (50 training, 48 testing) video footages collected from surveillance cameras that monitor pedestrian walkways around the UCSD campus. The dataset provides ground truth labels for two abnormal events: i) detection of non-pedestrian objects; ii) anomalous pedestrian motion patterns. We treat this dataset as the video data collected from the end devices since our end devices do not have cameras. We assume that each end device generates one sensor data stream. Each run of the experiment contains a total of 100 sensing cycles. Within each sensing cycle, the end devices are tasked to process a total of 8 seconds of video clips (with each video source sampled at 20 image frames per cycle). The computation task in the abnormal event detection application is defined below.

\textbf{Computation Tasks for Abnormal Event Detection:} performs motion feature extraction (optical flow) and object detection using Tiny YOLO \cite{redmon2016you}. 
\subsubsection{ Quality of Service (Application (Server) Side)}

 We perform similar experiments as those discussed in the previous case study. In particular, we evaluate all the schemes in terms of DHR and E2E delay. The results are shown in Figure~\ref{fig:abdelay} and Figure~\ref{fig:dhrab} respectively. We observe similar results of G-PATA as the previous case study. We also observe that both DHR and E2E delay results are worse than those in the previous section for all schemes. The reason is that i) the YOLO framework used in this application is more computationally intensive than the previous; ii) more image frames need to be processed per sensing cycle.

 \subsubsection{Edge Payoffs (end device Side)}

 The results of payoffs gained at the end devices are shown in Figure~\ref{fig:payoffab}. We observe that our scheme continues to provide significantly more amount of payoffs to the end devices than other baselines. G-PATA also outperforms TDA in terms of the payoff in this case study. This again demonstrates that G-PATA is more ``user-friendly" by allowing end devices to obtain more payoffs.

\subsection{\rv{ Overhead and Scalability Analysis}}

\begin{figure}[!tb]
	\centering
	\includegraphics[width=6.5cm]{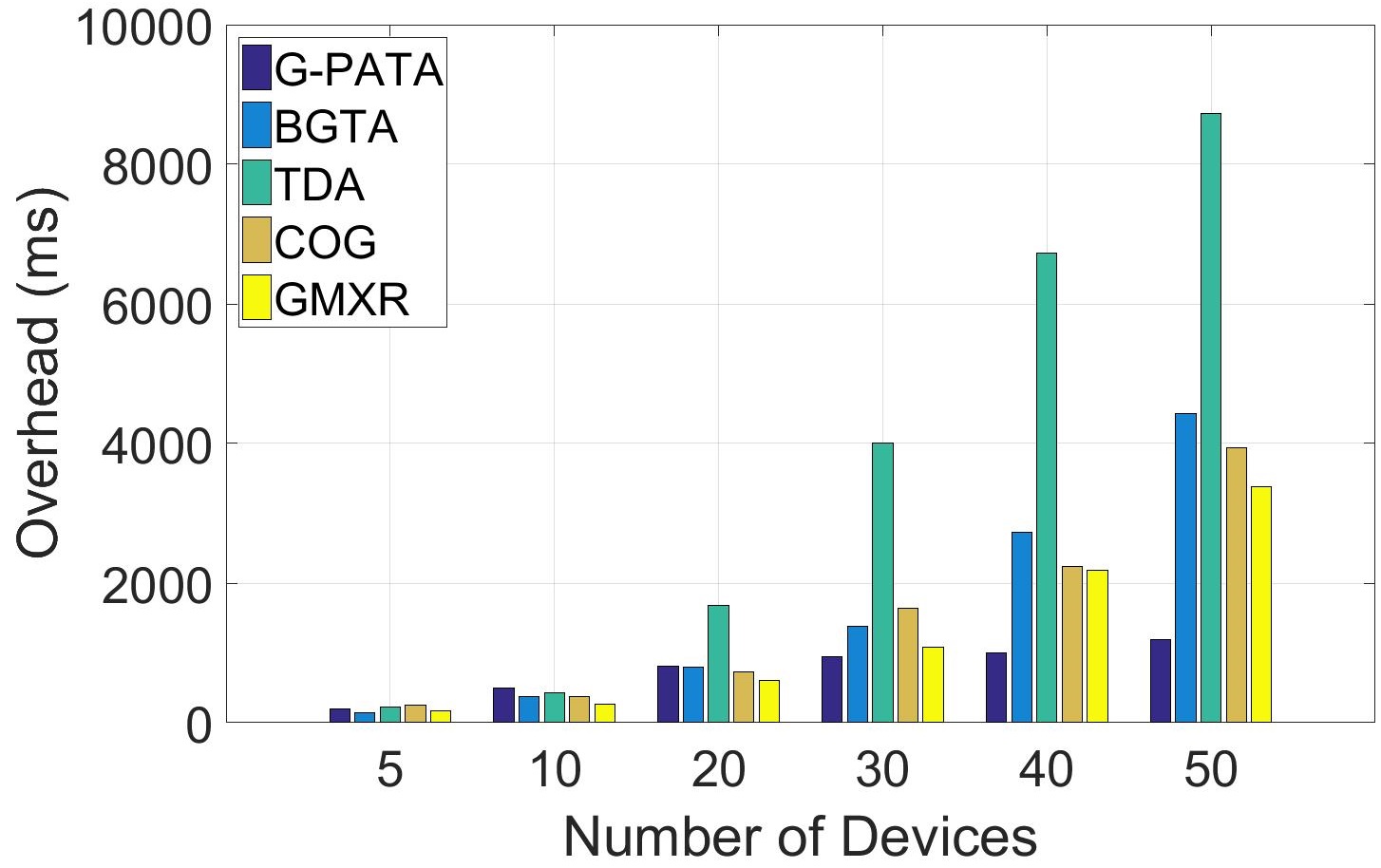}
	\vspace{-0.05in}
	\caption{Overhead of All Schemes vs \# of Devices}
	\label{fig:device}
	\vspace{-0.1in}
\end{figure}

\begin{figure}[!htb]
	\centering
	\includegraphics[width=6.5cm]{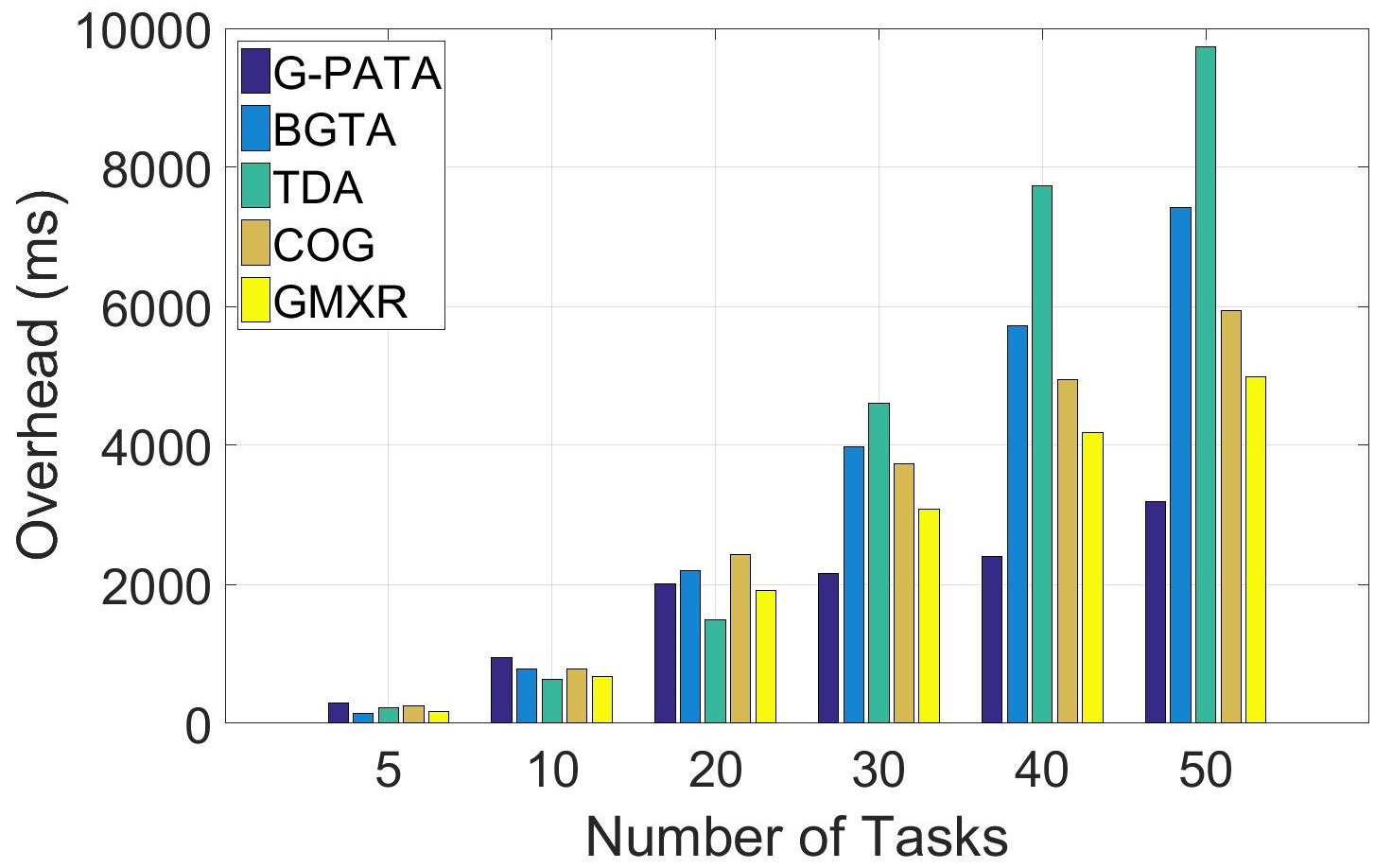}
	\vspace{-0.05in}
	\caption{Overhead of All Schemes vs \# of Tasks}
	\label{fig:task}
	\vspace{-0.1in}
\end{figure}

\begin{figure*}[!tb]
    \subfigure[][Number of Iterations]{
        \centering
        \includegraphics[width=0.30\textwidth]{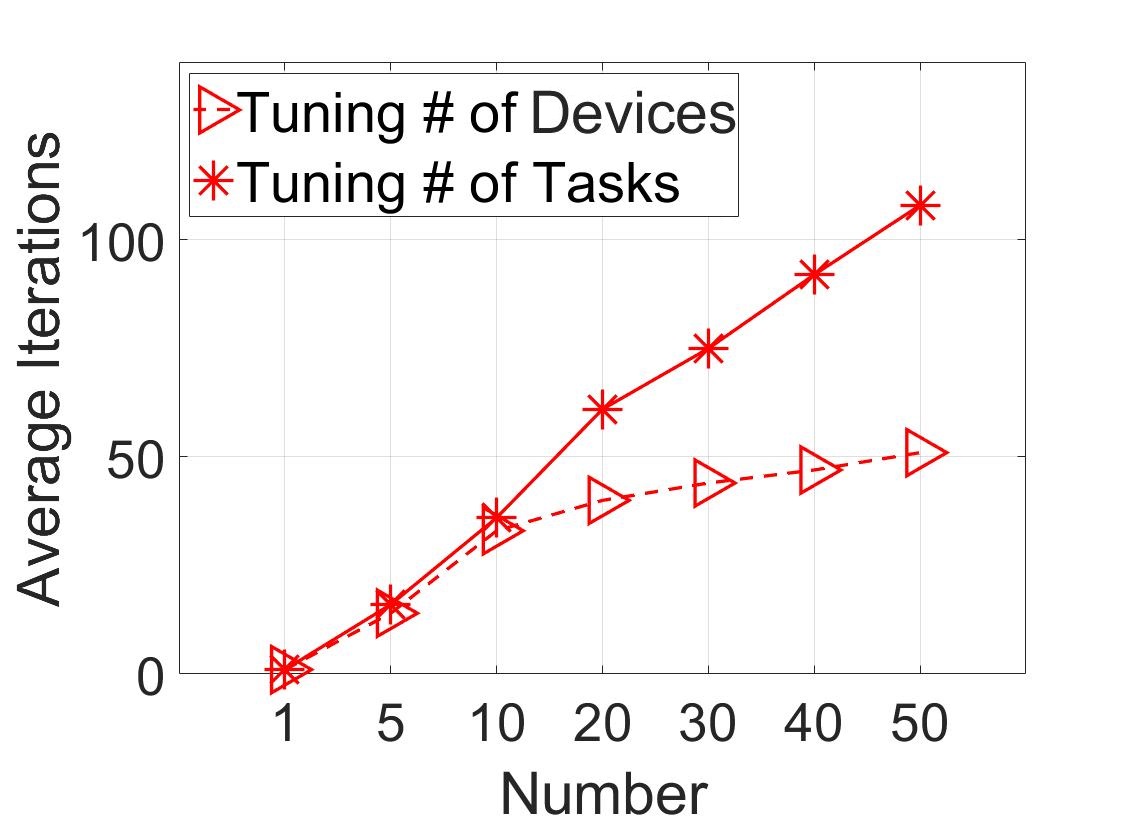}
        \vspace{-0.1in}
        \label{fig:conv}
    }
    \subfigure[][Overhead of G-PATA vs. \# of Devices]{
        \centering
        \includegraphics[width=0.30\textwidth]{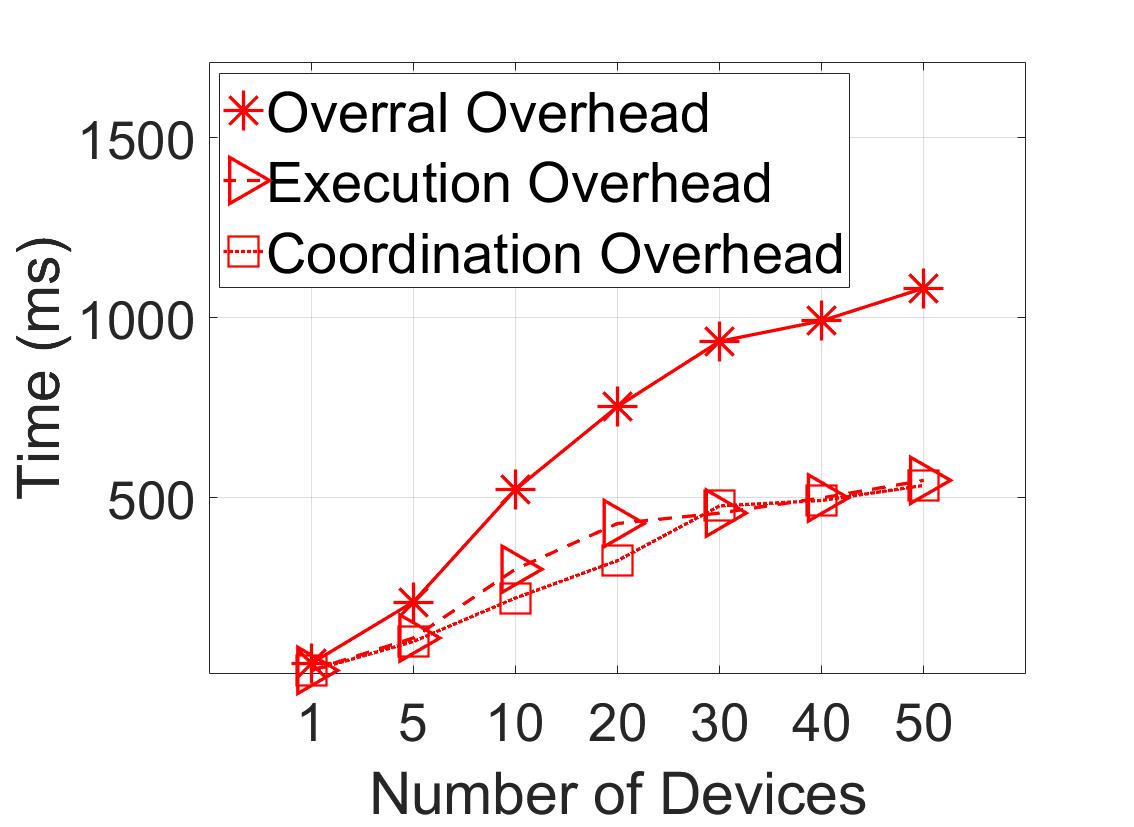}
        \vspace{-0.1in}
        \label{fig:ohd}
    }
        \subfigure[][Overhead of G-PATA vs. \# of Tasks]{
        \centering
        \includegraphics[width=0.30\textwidth]{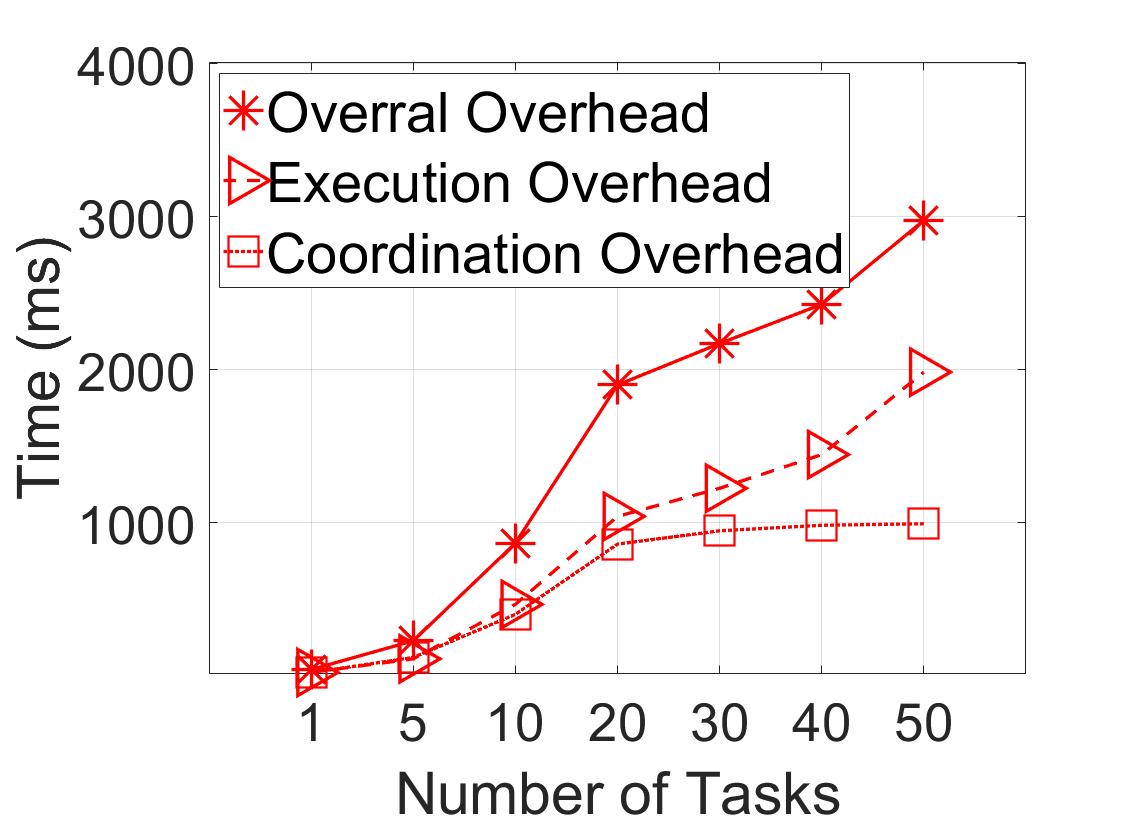}
        \vspace{-0.1in}
        \label{fig:oht}
    }
    \vspace{-0.1in}
    \caption{Scalability of G-PATA}
    \label{fig:scalability}
    \vspace{-0.15in}
\end{figure*}

\rv{In this subsection, we evaluate the algorithm overhead as well as the scalability of the compared schemes. In this set of experiments, we first tune the number of end devices while fixing the number of tasks. We then tune the number of tasks while fixing the number of devices. Such setting allows us to observe how the performance of each compared scheme changes when the workload increases.  We use the Collaborative Traffic Monitoring application for this experiment. We first present the algorithm overhead of each compared scheme in Figures 12 and 13. We exclude the random assignment baseline for the apparent reason that it always achieves the least overhead due to its simplicity. We can observe that G-PATA starts to outperform baselines when the number of devices or the number of tasks increases (e.g., $>$20). This demonstrates the scalability of the G-PATA. We also found the performance gain of G-PATA is more significant when increasing the number of devices as compared to increasing the number of tasks. This is due to the unique load balancing technique employed by G-PATA that ensures the algorithmic workload is evenly distributed across local edge groups. To take a closer look at the scalability of G-PATA, we further analyze the convergence and fine-grained algorithmic overhead of G-PATA when varying the number of devices/tasks in Figure 14. We can observe that the number of iterations it takes to converge, as well as the overall overhead of the G-PATA algorithm grows almost linearly as either the number of end devices or the number of tasks increases. We again observe that the number of tasks has a larger impact on the overhead of the G-PATA algorithm than the number of devices. Note that in the G-PATA implementation, we make sure the end devices are split into local groups so that G-PATA is only run on a few end devices as well as their assigned local edge server. This, along with the quasi-linear growth of the algorithm, ensure the overall overhead of the task allocation is manageable.}

	\section{Conclusion and Future Work}\label{sec:discuss}

This paper presents the G-PATA framework to solve the privacy-aware computation task allocation problem in the SSEC system. \dzf{ The G-PATA framework is motivated by a unique and practical observation that end users have diversified privacy requirements. In G-PATA, we develop novel privacy-aware bottom-up game theoretic framework approaches (i.e., ITAG and DPFP) that not only optimize the QoS of the application but also ensure the privacy requirements of end users are well satisfied. A new uncertainty-aware load balancing framework is also developed to efficiently balance the load of the G-PATA framework when only partial information of the end devices is available. We implemented the G-PATA framework on a real-world platform that consists of Nvidia Jetson TK1, TX1, and Raspberry Pi3 boards. The evaluation results from two real-world social sensing applications demonstrate that G-PATA can well protect user defined privacy while achieving significant performance gain in terms of reducing delay and deadline miss rate compared to sate-of-the-art baselines.}

Our work has some limitations and can be extended in future research. \rv{ One limitation lies in addressing security concerns. In particular,  we assume that end devices are not malicious. However, there may be malicious end devices that intentionally generate wrong results for tasks.  This issue can be addressed by combining G-PATA with verifiable computing techniques where the system maintains verifiable results by requiring a ``proof" of the correct execution of tasks~\cite{gennaro2010non}. For example, a peer verification mechanism was proposed in~\cite{vance2019towards} where the server can duplicate the computation to a judicious selection set of end devices (with incentives) and detect malicious attackers by comparing their computation results.} Also on the server side, we assume the server is honest-and-curious, i.e., it will not falsify the number of tasks for carrying out malicious attacks other than aggregating user information. In a practical setting, we can extend G-PATA with recent techniques such as honeyclients \cite{qassrawi2011detecting} that allows users to proactively detect malicious server behavior to further enhance the protection against malicious servers.

\rv{Secondly, as mentioned in Section III-C, G-PATA does not explicitly consider the potential privacy risks that are introduced by examining or linking the raw sensing data provided by end devices. We can extend G-PATA by integrating it with existing data privacy protection techniques. For example, various data obfuscation techniques have been proposed to sanitize imagery data and effectively conceal sensitive information in the images \cite{korshunov2013using, fan2019practical}. To address the privacy risk related to the data linkage, we can leverage techniques such as k-Unlinkability \cite{malin2008k} to make the raw data unlinkable. In future work, we plan to further investigate the trade-off between the QoS of the application and the user privacy protection when applying these data privacy mitigation techniques.}

Moreover, G-PATA assumes the end devices follow the game protocol honestly and only report truthful information. For example, if a device is in City A, it should not claim to be in City B. Alternatively, if a device has available CPU of 10\%, it should not claim to have 80-90\% available CPU. We can enforce such an assumption by ensuring that a client program of G-PATA installed on the end devices is responsible for identifying and reporting the information of devices.  The end users configure the privacy settings through a GUI and cannot intentionally modify the reported information. The privacy setting from user input is provided to the client program to properly mask the reported information (e.g., hide certain information that the user is not willing to provide). Therefore, we only need to ensure the client program is working properly and reports the truthful information.   In cases when end users do intend to hack the program and report incorrect information to gain higher rewards, G-PATA can perform software integrity protection \cite{chen2002oblivious,suh2003efficient} to ensure the client program has not been tampered with.

Finally, G-PATA does not consider the dynamic churn of end users.  In fact, we assume the availability of devices does not change within each sensing cycle. In practice, the availability of users could change at any point in time, which may lead to unfinished or partially finished tasks \cite{vance2019towards}.  G-PATA can be extended to address user churn by allocating important tasks to more than one end device so the task will be finished unless all devices associated with the task decide to quit simultaneously. Alternatively, G-PATA can also impose a high penalty cost for the devices who quit without finishing their tasks. \zy{G-PATA also did not consider the dynamic admission of new users into the system.  The newly participating devices may not be able to join the game-theoretic task right away until the current users finish the iterations.  We will extend the DPFP scheme to be able to admit new users within the iterations of the DPFP algorithm in the future.}

\section*{Acknowledgment}		
This research is supported in part by the National Science Foundation under Grant No. CNS-1845639, CNS-1831669, Army Research Office under Grant W911NF-17-1-0409. The views and conclusions contained in this document are those of the authors and should not be interpreted as representing the official policies, either expressed or implied, of the Army Research Office or the U.S. Government. The U.S. Government is authorized to reproduce and distribute reprints for Government purposes notwithstanding any copyright notation here on.

\bibliographystyle{IEEEtran}
\bibliography{main}

\vskip -1\baselineskip plus -1fil  \begin{IEEEbiography}[{\includegraphics[width=1in,height=1.25in,clip]{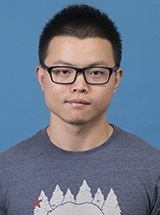}}]{Daniel (Yue) Zhang} obtained his Ph.D. from the department of Computer Science and Engineering at University of Notre Dame, Indiana in 2020. He received his M.S. degree from Purdue University, West Lafayette, Indiana, USA, in 2012 and a B.S. degree from Shanghai Jiao Tong University, Shanghai, China, in 2008. His research interests include human-centric computing, social sensing based edge computing ecosystem, truth analysis on social media, and cyber-physical systems. He is a student member of IEEE.
\end{IEEEbiography}\vskip -2\baselineskip plus -1fil 
 \begin{IEEEbiography}[{\includegraphics[width=1in,height=1.25in,clip]{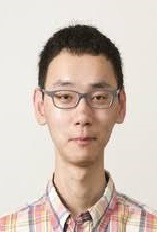}}] {Yue Ma} obtained his Ph.D. from the department of Computer Science and
Engineering at University of Notre Dame, Indiana in 2019.
He received his B.S. degree from Chengdu University of Technology, China, M.S. from University of
Electronic Science and Technology of China, China.
His research interests include real-time embedded
systems, reliable system design, power efficiency and
temperature-aware resource management.
\end{IEEEbiography}\vskip -2\baselineskip plus -1fil 
 \begin{IEEEbiography}[{\includegraphics[width=1in,height=1.25in,clip]{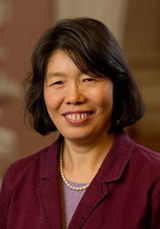}}] {Xiaobo Sharon Hu}
	(S'85-M'89-SM'02-F'16) received her B.S. degree from Tianjin University, China, M.S. from Polytechnic Institute of New York, and Ph.D. from Purdue University, West Lafayette, Indiana. She is Professor in the department of Computer Science and Engineering at University of Notre Dame. Her research interests include computing with beyond-CMOS technologies, low-power system design and cyber-physical systems. She has published more than 300 peer-reviewed papers in these areas. She  served as Associate Editor for IEEE Transactions on VLSI, ACM Transactions on Design Automation of Electronic Systems, and ACM Transactions on Embedded Computing, and is currently an associate editor for IEEE Transactions on CAD and ACM Transactions on Cyber-physical Systems  She was the General chair of 2018 Design Automation Conference (DAC), and Program Chair and TPC chair of 2016 and 2015 DAC. She received the NSF CAREER Award in 1997, and the Best Paper Award from Design Automation Conference in 2001 and ACM/IEEE International Symposium on Low Power Electronics and Design in 2018, etc.
\end{IEEEbiography}\vskip -2\baselineskip plus -1fil 
\begin{IEEEbiography}[{\includegraphics[width=1in,height=1.25in,clip,keepaspectratio]{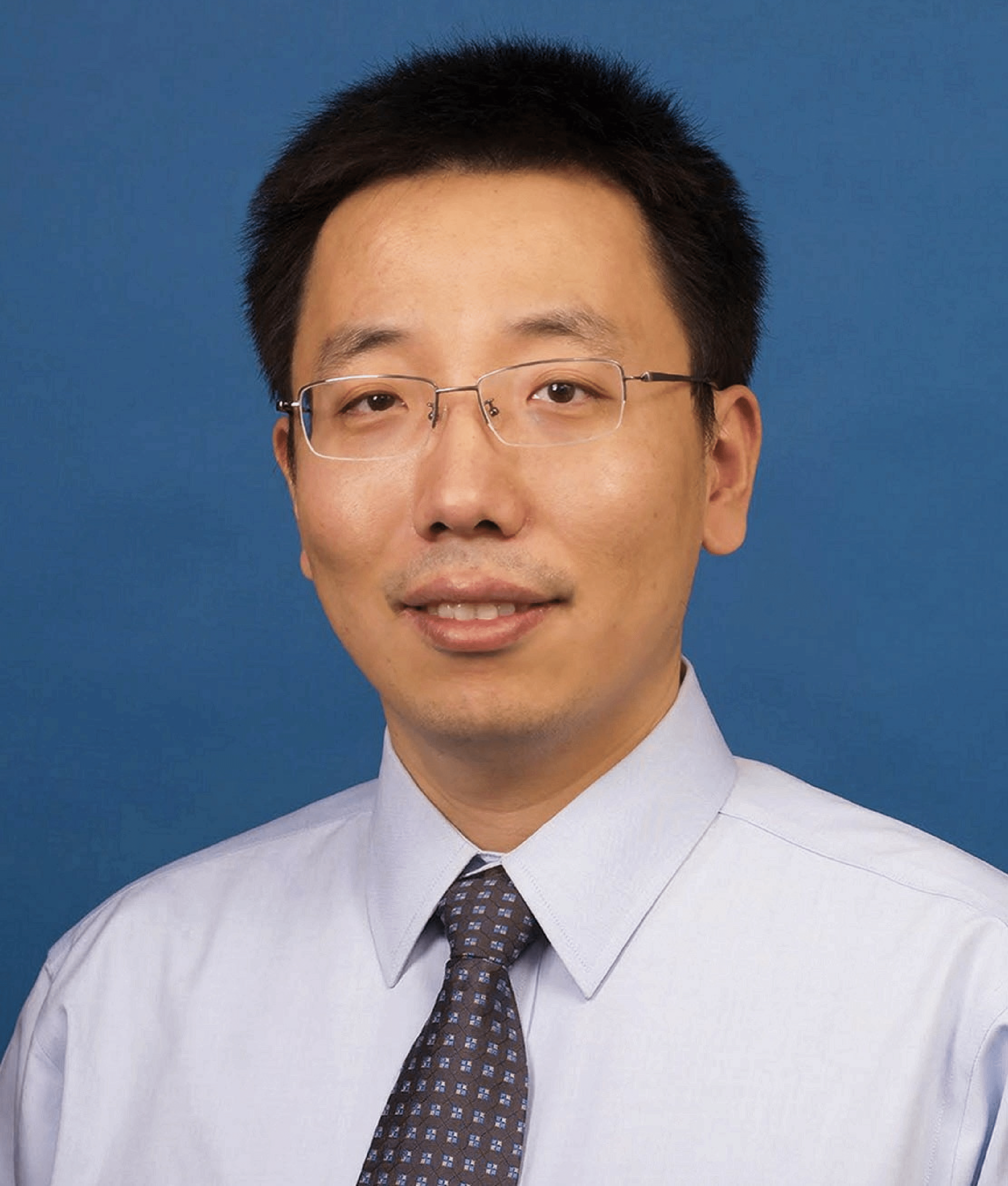}}]{Dong Wang} received his Ph.D. in Computer Science from University of Illinois at Urbana Champaign (UIUC). He is now an associate professor in the Department of Computer Science and Engineering at the University of Notre Dame. Dr. Wang's research interests lie in the area of social sensing, human-cyber-physical computing, edge computing, and smart cities applications. He has published more than 100 peer-reviewed papers in these areas. He received the NSF CAREER Award, Google Faculty Research Award, Army Research Office Young Investigator Program (YIP) Award, Wing-Kai Cheng Fellowship from the University of Illinois and the Best Paper Award of IEEE Real-Time and Embedded Technology and Applications Symposium (RTAS). He is a member of IEEE and ACM.
\end{IEEEbiography}

\end{document}